\documentclass[%
 reprint,
floatfix,
superscriptaddress,
showpacs,preprintnumbers,
 amsmath,amssymb,
 aps,
prl,
]{revtex4-2}

\usepackage[colorlinks]{hyperref}
\hypersetup{colorlinks = true,
linkcolor = blue,
filecolor = blue,
urlcolor = blue,
citecolor = blue}
\DeclareUnicodeCharacter{2212}{-}
\usepackage{xcolor}
\usepackage{times}
\usepackage{amssymb}
\usepackage{dsfont}
\usepackage{amsmath}
\usepackage{graphicx}
\usepackage{dcolumn}
\usepackage{bm}
\usepackage{epstopdf}
\usepackage{orcidlink}

\usepackage{ulem}

\usepackage{changes}

\definechangesauthor[name={}, color=green]{ep}

\begin{document}

\title{Fundamental limitations of single-particle Green's-function zeroes as probes of many-body topology}

\author{Emile~Pangburn}
\email{emile.pangburn@ipht.fr}
\affiliation{Institut de Physique Théorique, Université Paris Saclay, CEA CNRS, Orme des Merisiers, 91190 Gif-sur-Yvette Cedex, France}
\affiliation{Center for Computational Quantum Physics, Flatiron Institute, 162 5th Avenue, New York, New York 10010, USA}

\author{Olivier~Gingras\,\orcidlink{0000-0003-3970-6273}}
\email{ogingras@flatironinstitute.org}
\affiliation{Center for Computational Quantum Physics, Flatiron Institute, 162 5th Avenue, New York, New York 10010, USA}
\affiliation{Institut de Physique Théorique, Université Paris Saclay, CEA CNRS, Orme des Merisiers, 91190 Gif-sur-Yvette Cedex, France}

\begin{abstract}

We show that topological invariants constructed from single-particle Green's functions (GFs) cannot reliably diagnose the topology of interacting many-body states. Using coupled interacting SSH chains as a minimal example, we demonstrate that a spin-spin interaction can trivialize the many-body ground state without affecting the GF topological invariant. This breakdown originates from the GF's inability to probe electronic excitations in the Fock sectors responsible for the topological degeneracy. Consequently, GF zeroes are not associated with physical topological quasiparticles and cannot generally characterize interacting topological phases.

\end{abstract}

\maketitle

\textit{Introduction.---} Topological classification has been remarkably successful in non-interacting systems because the ground state reduces to a Slater determinant characterized by a single-particle wavefunction. As a result, topological properties are encoded in band structure quantities such as Berry phases, Chern numbers and winding numbers, efficiently computable from Bloch Hamiltonians~\cite{alexandradinata2014wilson, chiu2016classification}. This framework provides a complete and predictive classification of gapped phases based on symmetry and dimensionality, where bulk-boundary correspondence emerges naturally as non-trivial bulk invariants that guarantee robust edge or surface states~\cite{chen2011classification, pollmann2012symmetry, senthil2015symmetry, verresen2017one}.

In interacting systems, this success breaks down because the many-body ground state is no longer a Slater determinant. While some classifications survive, their practical diagnosis is far more difficult, typically requiring access to non-local quantities such as entanglement spectrum, edge degeneracies or high-order response functions.
For this reason, topological invariants based on single-particle Green's function (GF) denoted $\mathcal{G}$, which can be applied to interacting system, have been proposed~\cite{wang2010topological,wang2012simplified,gurarie2011single,manmana2012topological,yoshida2014characterization,iraola2021towards,lessnich2021elementary,blason2023unified,bollmann2024topological,peralta2023connecting,wagner2023mott,wagner2024edge,setty2024electronic,setty2024symmetry,nourafkan2013electric,dionne2025characterizing,konig2026topological,kogel2026poles,mitra2026signatures}.
An extensive review of GF topological invariants with examples of its application to the Hubbard dot and dimer is provided in the Supplemental Material (SM).

In the non-interacting limit, these GF invariants correspond exactly to the band structure ones. Therefore, single-particle GF poles provide a quasiparticle interpretation of the topology of the non-interacting limit. 
Including interactions, the single-particle GF can vanish because of spectral-weight redistribution and the resulting zeroes also contribute to the GF topological invariants. This raises the question of the meaning of non-trivial topology originating from GF zeroes.

A zero of the GF is defined by a frequency $\omega_z$ at which
\begin{equation}
    \label{eq:GF_zero}
    | \det \mathcal{G}(\omega = \omega_z) | = 0.
\end{equation}
Green's function zeroes were first invoked as a solution to the Luttinger theorem violation~\cite{dzyaloshinskii2003some, stanescu2007theory,rosch2007breakdown, seki2017topological, skolimowski2022luttinger, dave2013absence}. Because this theorem formally counts both poles and zeroes, it was realized that the Luttinger theorem could be saved for a Mott insulator at half-filling by counting the Luttinger surface: the contribution from the zeroes. Extending this concept away from half-filling, these surfaces were found to be a ubiquitous feature of correlated metals and it was argue they could provide a way to understand the pseudogap phase~\cite{stanescu2006fermi, sakai2009evolution, sakai2010doped, yang2006phenomenological, scheurer2018topological, worm2024fermi}. 
For these reasons, a compelling quasiparticle interpretation of GF zeroes was given in terms of emergent spinons, charge neutral excitations~\cite{fabrizio2022emergent, fabrizio2023spin, blason2024luttinger, pasqua2025fermi}. 
Because of their vanishing charge, they would not contribute to electrical conductivity, but could contribute to thermal conductivity. If true, this provides a practical correspondence between single-particle GF and many-body phenomena.

Followed up the argument that this quasiparticle interpretation could also apply to cases where GF zeroes are gapped with no Luttinger surface~\cite{yoshida2014characterization, wagner2023mott, wagner2024edge}. A bulk-boundary correspondence was established and linked to spin physics. This theory is attractive as it leads to a direct physical meaning of topological invariants based on GF zeroes, where GF topology would directly capture many-body topology. Practical computations in some explicit model are suggesting these invariants are useful for predicting topology and physical properties~\cite{yoshida2014characterization, wagner2023mott}.

So far, only few papers have raised doubt about the usefulness of these GF invariants~\cite{manmana2012topological,he2016topological1,he2016topological2,zhao2023failure,aligia2025probing,markov2026luttinger}. In particular, Ref.~\citenum{zhao2023failure} highlighted an unphysical dependence of GF based topological invariants on the position of the chemical potential. To avoid this issue, we restrict our analysis to particle–hole symmetric systems.
In this manuscript, we demonstrates that these doubts are in fact correct by considering two topological chains, trivializing them with a many-body interaction, and showing the the GF invariants remain unchanged. We therefore confirm that GF invariants aren't guarantee to capture the full many-body topology of strongly interacting systems. We explain that this result is expected, given that GFs probe electronic excitations in only a few selected Fock sectors and generally miss those responsible for topological transitions. As a consequence, we show that GF zeroes have no direct physical meaning in terms of quasiparticle excitations These conclusions are independent of the dimensionality of the system considered.

\begin{figure}[t!]
    \centering
    \includegraphics[width=\linewidth]{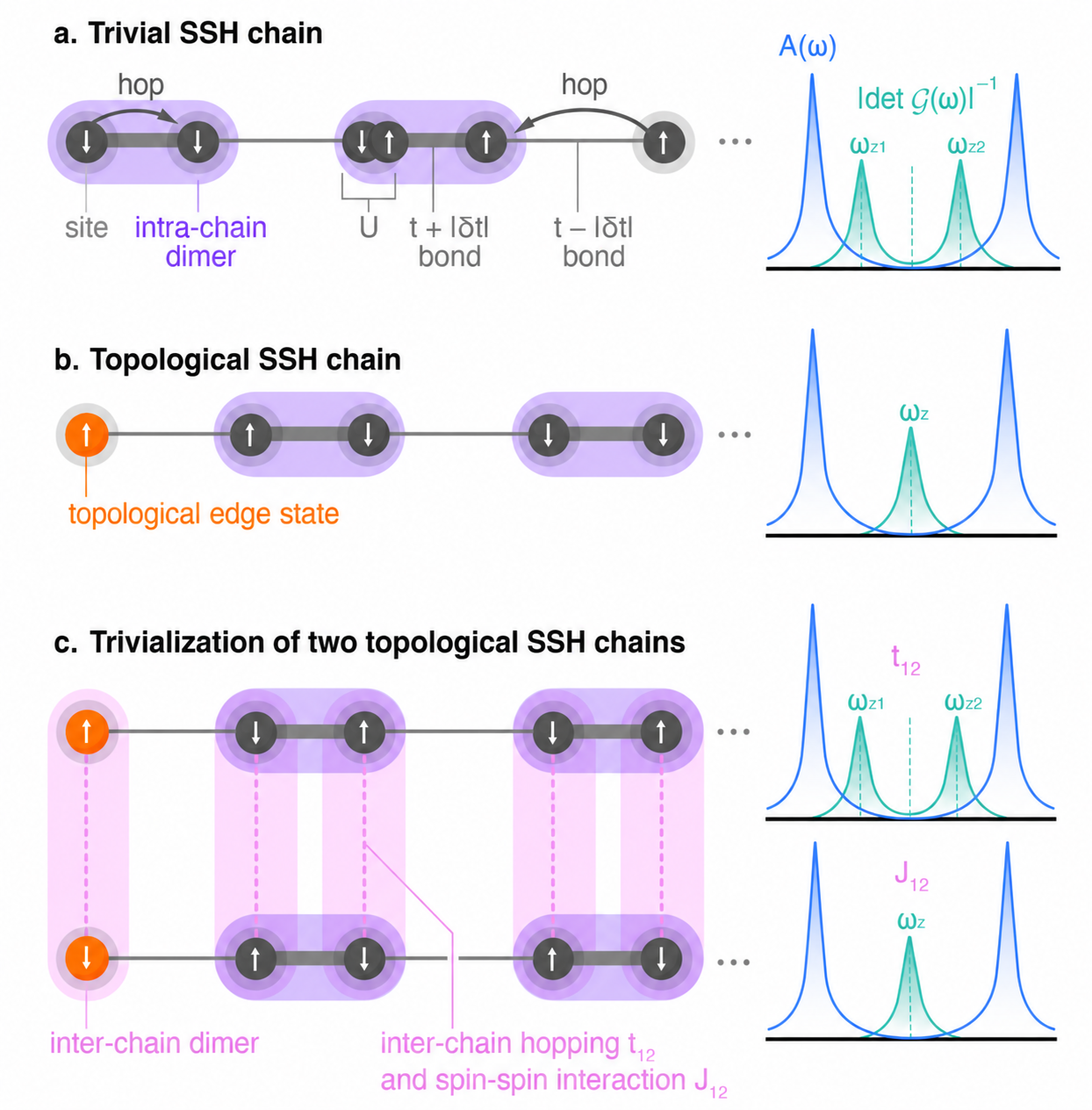}
    \caption{
    Trivialization of two topological chains.
    Left figures: schematics of the models considered.
    Right figures: spectral function $A(\omega)$ in blue and Green's function invariant $|\det \mathcal{G}(\omega)|^{-1}$ in green.
    (a)~Trivial configuration of the interacting SSH chain. The strong bond, with larger hopping $t+|\delta t|$, dimerizes the edge.
    (b)~Topological configuration of the interacting SSH chain. The edge is connected by a weak bond with hopping $t-|\delta t|$, which cannot dimerize and acts as an independent edge mode.
    (c)~Including a coupling between two topological chains allows them to trivialize by forming inter-chain dimers.
    }
    \label{fig:SSH_chains}
\end{figure}

\textit{Two coupled SSH chains.---} Throughout this work, we consider an extension of the Su–Schrieffer–Heeger (SSH) chain~\cite{su1979solitons} in the presence of a local Coulomb repulsion term with strength $U$, in order to investigate the relationship between GF zeroes and the topology of the interacting ground state. 
Such a chain with index $l$ is characterized by the Hamiltonian 
\begin{align}
    \label{eq:SSH_Ham}
    H_l \equiv \sum_{i\sigma} t_i\left[c^\dagger_{li\sigma}c_{li+1\sigma}+h.c.\right]
    + 
    U \sum_i n_{li\uparrow}'n_{li\downarrow}',
\end{align}
where $c^\dagger_{li\sigma}$ ($c_{li\sigma}$) creates (annihilates) an electron with spin $\sigma$ at site $i$ on the chain $l$, $t_i \equiv (t+(-1)^i\delta t)$ is the bond dependent hopping with $t$ the uniform hopping amplitude, $\delta t$ is the dimerization parameter and $(-1)^i$ alternates between bounds, and $n_{li\sigma}' \equiv c^\dagger_{li\sigma}c_{li\sigma}-\tfrac12$ enforces particle-hole symmetry.
A visual representation of this chain is shown in Fig.~\ref{fig:SSH_chains}(a,b).

As detailed in the SM, for an even number of sites and depending on the sign of $\delta t$ at the edges, such a chain can either host a trivial or topological ground state. If $\delta t$ increases the hopping at the edge, the resulting strong bond forms a trivial boundary dimer with no edge state (Fig.~\ref{fig:SSH_chains}(a)). If instead $\delta t$ decreases the hopping at the edge, the bond is weak and leaves an unpaired site at each boundary, equivalent to two Hubbard dot as topological edge states (Fig.~\ref{fig:SSH_chains}(b)). This topology is captured by GF zeroes. Under periodic boundary conditions (PBC), the zero spectrum is fully gapped in both the topological and trivial phases. In contrast, under open boundary conditions (OBC), the zero spectrum becomes gapless in the topological phase while remaining gapped in the trivial one. This distinction reflects the doublet versus singlet nature of the edge states in the two distinct configurations. In these scenarios, there is indeed a one-to-one correspondence between GF zeroes and the underlying many-body topology, because it is dictated by a single-particle term of the Hamiltonian ($\delta t$).

We now consider two such chains in their topological configuration, with labels $l=1,2$, and we couple them as depicted in Fig.~\ref{fig:SSH_chains}(c). We use two types of couplings: a one-body inter-chain hopping and a two-body spin-spin interaction, with amplitudes $t_{12}$ and $J_{12}$, respectively. They are expressed as
\begin{align}
    H'_{12} = 
    J_{12}\sum\limits_i \vec{S}_{1i}\cdot\vec{S}_{2i} 
    + 
    t_{12}\sum\limits_{i\sigma}\left[c^\dagger_{1i\sigma}c_{2i\sigma}+h.c.\right], 
\end{align}
where $\vec{S}_{li} \equiv c^\dagger_{li\tau}\vec{\sigma}_{\tau\tau'}c_{li\tau'}$ is a local spin operator with $\vec{\sigma}$ the vector of Pauli matrices.
The full Hamiltonian is $H_{12} \equiv H_1 + H_2 + H'_{12}$. A finite $t_{12}$ couples the chains at the single-particle level which breaks the sublattice symmetry. In contrast, a finite $J_{12}$ introduces a purely many-body interaction that preserves all symmetries.

\begin{figure}[b!]
    \centering
    \includegraphics[width=\linewidth]{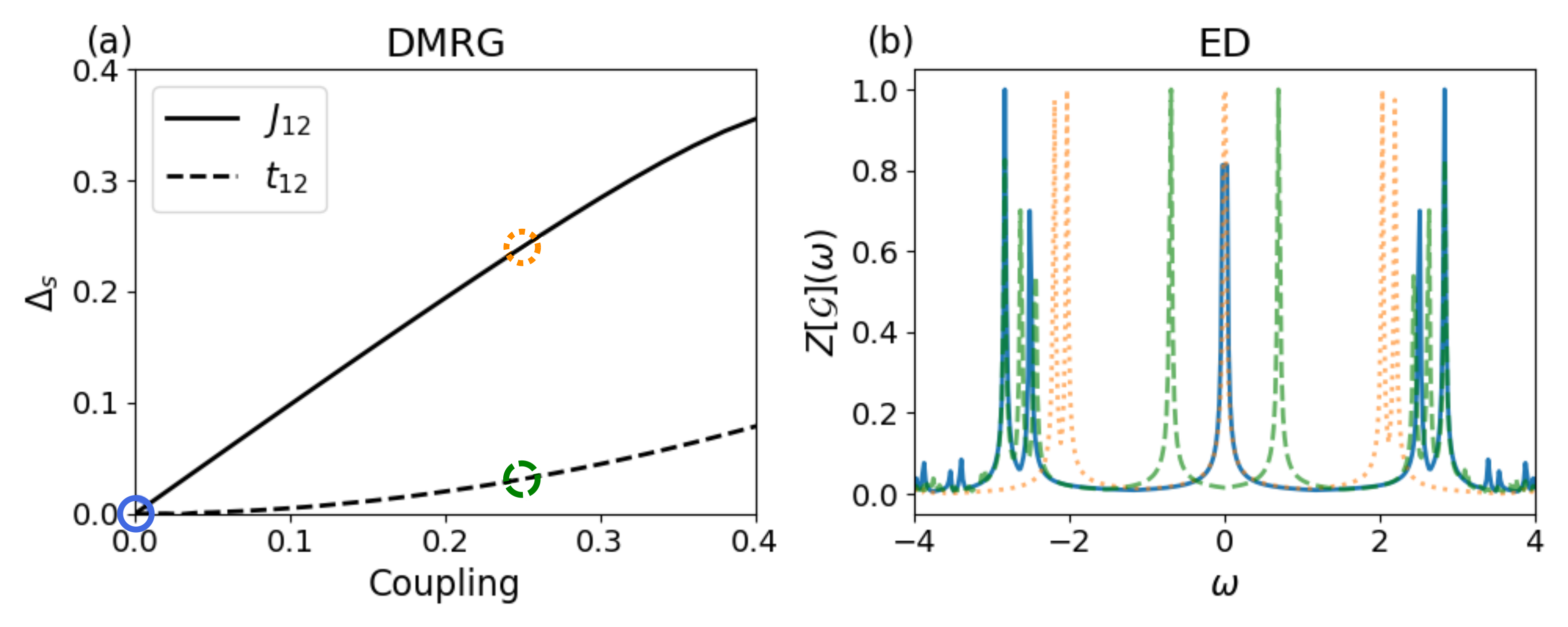}
    \caption{
        Inaccurate topology inferred from Green's function zeroes in two SSH chains coupled by many-body interactions. We study the model of Eq.~(\ref{eq:SSH_Ham}) using the parameters $t=\tfrac34$, $\delta t = \tfrac t3$ and $U=8$.
        (a)~Spin gap for two interacting SSH chains of length $L=24$ as a function of inter-chain Heisenberg coupling $J_{12}$ and hybridization $t_{12}$, computed with DMRG.
        (b)~GF zeroes probe $Z[\mathcal{G}](\omega)$ computed with ED for two interacting SSH chains of length $L=6$ coupled with $(J_{12},t_{12})=(0,0)$ (full blue line), $(J_{12},t_{12})=(\tfrac14, 0)$ (dotted orange line) and $(J_{12}, t_{12})= (0,\tfrac14)$ (dashed green line), with open boundary conditions. Although a finite $J_{12}$ trivializes the two chains, the Green's function invariant remains unchanged.
    }
    \label{fig:spin_gap_vs_GFz}
\end{figure}

\textit{Trivialization through coupling.---}
In order to demonstrate that GF zeroes do not reliably characterize the topology of the many-body state in the bulk, in particular when influenced by many-body interactions, we want to compare the GF invariant based on the zeroes, $|\det \mathcal{G}(\omega)|^{-1}$, with observables associated to the ground-state topology: the spin and the charge gaps.
For numerical stability, instead of computing the determinant of the GF, we compute the eigenvalue spectrum $\lambda_\mathcal{G}$ of $\mathcal{G}$. The zeroes correspond to vanishing eigenvalues, and we plot the regularized quantity $Z[\mathcal{G}](\omega)\equiv\left[\min |\lambda_{\mathcal{G}}(\omega)|\right]^{-1}$ to highlight the zeroes. The poles are not analyzed here since their topology is trivial when $U\ne 0$. 
Another critical fingerprint of symmetry protected topological states (the only topological phases possible in one dimension) that we verify is that the ground state is non-degenerate under PBC and degenerate under OBC, characteristic of a gap in the bulk and modes at the edges~\cite{chen2011classification} (See SM).

In Fig.~\ref{fig:spin_gap_vs_GFz}(a), we present the effect of coupling the two topological interacting SSH chains on an indicator of topology: the spin gap $\Delta_s$ under OBC, obtain using the density matrix renormalization group (DMRG, see SM). The simulated chains have $24$ sites each, and their associated interacting SSH Hamiltonian Eq.~(\ref{eq:SSH_Ham}) employs the following parameters: $U=8$, $t=\tfrac34$ and $\delta t =\tfrac14$. We find that any finite $J_{12}$ or $t_{12}$ opens a spin gap, reflecting the formation of singlets between the two topological edge spin-\textonehalf{} modes and therefore signaling the trivialization of the system as soon as inter-chain coupling is introduced.

On the other hand, in Fig.~\ref{fig:spin_gap_vs_GFz}(b), we present the GF topological indicator $Z[\mathcal{G}]$ for three cases: in full blue, dotted orange and dashed green, $(J_{12}, t_{12})=(0,0)$, $(\tfrac14,0)$ and $(0,\tfrac14)$, respectively. We use exact diagonalization (ED, see SM) since we need the full spectrum to compute $\mathcal{G}$, and although the size of the chains is now $L=6$, the parameters of Eq.~(\ref{eq:SSH_Ham}) are the same as in the other panel.
When both couplings are absents, each SSH chain is in a topological phase and exhibits GF zeroes associated with boundary spin modes under OBC, as discussed above.
When the inter-chain spin-spin interaction is  turned on ($J_{12}\ne 0$), these boundary GF zeroes remain robust, despite the fact that the boundary spins are gapped by singlet formation. In contrast to the many-body coupling $J_{12}$, when the topology is trivialized with a finite one-body coupling $t_{12}$, the GF zeroes succeed in capturing the trivialization.

These results demonstrate that GF zeroes are generally insensitive to topological changes induced by many-body interactions. They can encode single-particle physics but do not provide a reliable diagnostic of interacting topological phases, either at the boundary or in the bulk. 
To understand this short-coming and determine whether GF zeroes are capable of probing topological edge states and capturing the relevant topological quasiparticle excitations, one must analyze both the nature of these modes and whether GF zeroes couple to them.

The topological nature (and classification) of these edge states is tied to the anti-unitary sublattice (particle-hole) symmetry. A single chain can host a spinful topological edge mode protected by this symmetry~\cite{verresen2017one}.
When Hubbard interactions are included, charge fluctuations are suppressed and the boundary modes reduce to effective spin-\textonehalf{} modes. These can combine into either nontrivial half-integer or trivial integer spin representations~\cite{chen2011classification, pollmann2012symmetry}.
Therefore, in the absence of inter-chain interactions, an infinite set of SSH chains with an even number of sites can have any integer number of protected boundary modes, described by a $\mathds{Z}$ classification.
In the presence of inter-chain quartic interactions, $4$ modes can interact and be completely gapped~\cite{fidkowski2010effects,queiroz2016dimensional}, reducing the classification to the $(\mathds{Z}\mod4)$ classification.

The reason why the single-particle electronic GF fails to capture the properties of the edge modes in the coupled SSH chains is that it is not detecting the spin physics described in the previous paragraph. More precisely, a spin-up electronic creation operator $c^\dagger_\uparrow$ connect Fock sectors $(N_{\uparrow},N_{\downarrow})$ and $(N_{\uparrow}+1,N_{\downarrow})$, and similarly for a spin-down operator. However, in a SSH chain with $U\ne 0$, the low-energy edge excitations do not carry charge. The edge degrees of freedom studied here are purely spin-like, such that the relevant operators are spin-flips such as $S^+ = c^\dagger_{\uparrow}c_{\downarrow}$,
which connects the Fock sectors $(N_{\uparrow},N_{\downarrow})$ with $(N_{\uparrow}+1,N_{\downarrow}-1)$. The Fock sectors probed by single-particle GF are then not always overlapping with those responsible for the many-body topological degeneracy such that GF zeros can only partially probe the topology of the system.
This intuition can be made quantitative through a quasiparticle interpretation of GF zeroes~\cite{fabrizio2022emergent, fabrizio2023spin, wagner2023mott, wagner2024edge}, which clarifies which types of excitations GF zeroes are actually sensitive to.

\textit{Quasiparticle interpretation of GF zeroes.---}
The idea to treat zeroes as quasiparticles is rooted in the fact that, for an isolated atom with Coulomb repulsion $U$, the self-energy $\Sigma$ can be represented with a single pole. Therefore, in the deep Mott regime on the lattice and at half-filling, one can express the self-energy using a continued-fraction expansion as~\cite{wagner2023mott}
\begin{equation}
    \label{eq:Mott_self}
    \Sigma(\textbf{k}, \omega) \approx \frac{U^2/4}{\omega-\tilde H_0(\textbf{k)}},
\end{equation}
where $\tilde H_0(\textbf{k})$ is an effective non-interacting Hamiltonian that describes the dispersion of the zeroes. This effective Hamiltonian preserves the symmetries of the original Hamiltonian $H_0$, but with renormalized parameters. It was shown in Ref.~\citenum{pangburn2025topological} that one obtains an identical structure within the Hubbard operator (HO) method, detailed and benchmarked in the SM. We note that away from half-filling, the structure of the zeroes is more complicated and one can no longer express $\Sigma(\mathbf{k},\omega)$ in this simple form. Instead, $\tilde{H_0}$ would then behave as an interacting Hamiltonian~\cite{sakai2009evolution, sakai2010doped}.

Given a self-energy with a pole structure such as in Eq.~(\ref{eq:Mott_self}), it was proposed in Ref.~\citenum{fabrizio2022emergent} that, because these poles (GF zeroes) have a quasiparticle-like dispersion $\tilde H_0(\textbf{k})$, they could be interpreted as quasiparticles such as spinon-like excitations. It suggests that the topology of fractionalized excitations could be accessed through an effective single-particle Hamiltonian ($\tilde{H}_0$). However, we argue that this is a leap of faith, as the GF zeroes do not correspond to emergent quasiparticles but rather to signatures of the breakdown of the electron as an elementary excitation, a phenomenon recognized early~\cite{weinberg1965evidence}. In fact, for the coupled SSH chains with $J_{12}\neq 0$, the GF zero spectrum remains gapless while the ground state is fully gapped, highlighting the decoupling between GF zeroes and physical excitations. Therefore, while the gaplessness of GF zero spectrum is protected by a topological invariant, the conclusion from our work is that this invariant is unable to reliably capture the underlying many-body topology. We now show this in detail.

\textit{Topological invariant of the zeroes.---}
A one-dimensional bulk-gapped chiral-symmetric single-particle Hamiltonian $H_0$ such as the one studied in the work can be continuously deformed and spectrally flattened into an off-diagonal form, allowing the computation of a winding number $\nu[H_0]$ that classifies the system~\cite{chiu2016classification}. This applies to the zeroes Hamiltonian $\tilde H_{0}(\textbf{k})$, since it is a renormalized version of the original $H_0$ and therefore the topology from GF zeroes is protected by sublattice symmetry. In the case where the winding number of $\tilde H_{0}$ is non-trivial, in OBC, GF zeroes edge modes will be pinned at $\omega=0$. The only way to shift the zero modes away from zero energy without closing the gap of zeroes is to explicitly break the sublattice symmetry. We find that for the single chain, the interacting topological invariant computed from the winding of $\tilde{H}_0(\mathbf{k})$ remains identical to its non-interacting value ($U=0$) for all interaction strengths considered (see SM). 

In particular for the two interacting SSH chains, we investigate the impact of inter-layer coupling $J_{12}$ on the structure of the zeroes Hamiltonian obtained using the HO method. Labeling each chain by $l=1,2$, we also break their bulk into two sublattices: A$_l$ and B$_l$. In the basis $\{$A$_1, $A$_2, $B$_1, $B$_2\}$ and in the absence of inter-chain coupling, the zeroes Hamiltonian takes the form
\begin{align}
    \tilde H_{0}(\textbf{k}) = 
    \begin{pmatrix}
        0 & \tilde H_{0}^{\text{AB}}(\textbf{k})\\
        \tilde H_{0}^{\text{AB}}(\textbf{k})^\dagger  & 0\\
    \end{pmatrix}.
\end{align}
As detailed in the SM, $\tilde{H}_{0}^{\text{AB}}(\textbf{k})$ is expressed as $\text{diag} \left[q_0(\textbf{k}), q_0(\textbf{k}) \right]$ where $q_0(\textbf{k})$ couples A$_l$ with B$_l$.
In this case, the winding number associated with the GF zeroes is simply given by twice the winding number of $q_0(\textbf{k})$: $\nu[\tilde{H}_0] = 2 \nu[q_0]$ (See SM).

We now turn on either $t_{12}$ or $J_{12}$. These terms couple the sublattices A$_1$ and \ A$_2$, along with the sublattices B$_1$ and B$_2$. $t_{12}$ explicitly breaks the sublattice symmetry, whereas $J_{12}$ preserves it. As a result, the zeroes Hamiltonian acquires a more general structure with non-vanishing diagonal parts, which can be calculated using the HO method (See SM), yielding
\begin{align}
    \nonumber
    \tilde{H}_0 
        & = \begin{pmatrix}
            \tilde{H}^{\text{AA}}_0 & \tilde{H}^{\text{AB}}_0\\
            \left(\tilde{H}_0^{\text{AB}} \right)^\dagger & \tilde{H}^{\text{BB}}_0\\
        \end{pmatrix}
        \ \text{with} \ \
        \tilde H_{0}^{\text{AA/BB}} =
        \begin{pmatrix}
            0 & q_{12}\\
            q_{12} & 0 \\
        \end{pmatrix},
    \\
    & \quad \text{with}  \  \ q_{12} =q_{12}[t_{12},J_{12}].
\end{align}

The effect of $\tilde{H}_0^{\text{AA/BB}} \ne 0$ is to split the zero modes that appear under OBC away from zero energy, therefore trivializing the system, in this case because of the breaking of the sublattice symmetry. As shown explicitly in the SM, the hopping-induced term $q_{12}[t_{12}\ne 0, J_{12}]$ is generically non-zero away from the infinite-temperature limit regardless of $J_{12}$, while $q_{12}[t_{12}=0,J_{12}\ne 0]$ vanishes identically when the particle number in each layer is conserved.
This implies that, considering only a spin–spin interchain coupling $J_{12}$ affects the GF zeroes only through a renormalization of $\tilde H_{0}^{\text{AB}}$, induced by the interaction. Therefore, they remains in the same topological class as in the decoupled case, even though the many-body phase itself is fully trivial. As a result, the topology inferred from GF zeroes is incorrect in this situation.

\textit{Generalization to 2D.---}
While the previous discussion focused on a one-dimensional system, the conclusions can easily be extended to higher dimensions. However, exact calculations of the spectrum become rapidly intractable and we have to employ two different methods to obtain the ground state and the spectral function. This is not ideal but nevertheless allows us to make general statements. To support the generality of the results to two dimensions, we investigated (See SM) a two-dimensional generalization of the SSH chain: the Benalcazar–Bernevig–Hughes (BBH) model~\cite{benalcazar2017quantized}, using auxiliary-field quantum Monte Carlo for the ground state and the HO method for the spectrum. This model realizes a higher-order topological insulating phase protected by chiral and crystalline symmetries. In the non-interacting limit, this model hosts zero-dimensional corner modes, which has the same nature as the SSH edge states~\cite{jiang2023mottness}, and including and on-site repulsion, the interacting BBH model exhibits a two-dimensional realization of the same pole–zero mechanism observed in the SSH chain, but with four independent localized boundary degrees of freedom at each corner. By coupling two of these BBH layers through an inter-layer spin-spin interaction, we again can show that the GF invariant is incapable of capturing the right many-body topology.

\textit{Conclusion.---}
The results presented here on the coupled interacting SSH chains and the coupled interacting BBH layers clarify the precise scope of single-particle Green’s function approaches to interacting topology. Green’s function zeroes successfully reproduce topological structures inherited from non-interacting band theory, but fail once topology is governed by intrinsically many-body processes. In particular, interaction-driven trivialization can occur without any corresponding change in the Green’s function zero topology.

This breakdown originates from a fundamental mismatch between the excitations probed by single-particle Green’s functions and those responsible for the many-body topological structure. In the Mott regime studied here, the relevant low-energy edge excitations are spin-like collective modes, while the electronic Green’s function probes charged excitations in different Fock sectors. As a result, Green’s function zeroes cannot generally serve as universal probes of interacting topological phases.

More broadly, our work shows that extending topological band-theory concepts to strongly correlated systems cannot, in general, rely solely on single-particle quantities. A faithful characterization of interacting topology likely requires observables directly sensitive to the relevant collective excitations or to the many-body ground-state structure itself. More specifically for Mott insulators where the low-energy excitations are spin-like, a more appropriate framework could be to extend the single-particle Green's function invariants to multi-particle Green's functions such as dynamical spin correlators.

\begin{acknowledgments}
    \textit{Acknowledgments.---}We thank Anurag Banerjee, Antoine Georges, Giorgio Sangiovanni, Andrew J. Millis, Miles Stoudenmire, John Zima and Shiwei Zhang for valuable discussions, along with Brandon K. Eskridge for their support regarding some of the calculations.
    The Flatiron Institute is a division of the Simons Foundation.
\end{acknowledgments}

\bibliography{Bib_zeroes_Spin.bib}

@article{hubbard1963electron,
    author = {Hubbard, J.},
    title = {Electron correlations in narrow energy bands},
    journal = {Proceedings of the Royal Society of London. A. Mathematical and Physical Sciences},
    volume = {276},
    number = {1365},
    pages = {238-257},
    year = {1963},
    month = {11},
    issn = {0080-4630},
    doi = {10.1098/rspa.1963.0204},
    url = {https://doi.org/10.1098/rspa.1963.0204},
}

@article{hubbard1964electron,
    author = {Hubbard, J.},
    title = {Electron correlations in narrow energy bands {III}. {A}n improved solution},
    journal = {Proceedings of the Royal Society of London. A. Mathematical and Physical Sciences},
    volume = {281},
    number = {1386},
    pages = {401-419},
    year = {1964},
    month = {09},
    issn = {0080-4630},
    doi = {10.1098/rspa.1964.0190},
    url = {https://doi.org/10.1098/rspa.1964.0190},
}

@article{weinberg1965evidence,
  title = {Evidence {T}hat the {D}euteron {I}s {N}ot an {E}lementary {P}article},
  author = {Weinberg, Steven},
  journal = {Phys. Rev.},
  volume = {137},
  issue = {3B},
  pages = {B672--B678},
  numpages = {0},
  year = {1965},
  month = {Feb},
  publisher = {American Physical Society},
  doi = {10.1103/PhysRev.137.B672},
  url = {https://link.aps.org/doi/10.1103/PhysRev.137.B672}
}

@article{roth1969electron,
  title = {Electron {C}orrelation in {N}arrow {E}nergy {B}ands. {I}. {T}he {T}wo-{P}ole {A}pproximation in a {N}arrow {$S$} {B}and},
  author = {Roth, Laura M.},
  journal = {Phys. Rev.},
  volume = {184},
  issue = {2},
  pages = {451--459},
  numpages = {0},
  year = {1969},
  month = {Aug},
  publisher = {American Physical Society},
  doi = {10.1103/PhysRev.184.451},
  url = {https://link.aps.org/doi/10.1103/PhysRev.184.451}
}

@article{su1979solitons,
  title = {Solitons in {P}olyacetylene},
  author = {Su, W. P. and Schrieffer, J. R. and Heeger, A. J.},
  journal = {Phys. Rev. Lett.},
  volume = {42},
  issue = {25},
  pages = {1698--1701},
  numpages = {0},
  year = {1979},
  month = {Jun},
  publisher = {American Physical Society},
  doi = {10.1103/PhysRevLett.42.1698},
  url = {https://link.aps.org/doi/10.1103/PhysRevLett.42.1698}
}

@article{berry1984quantal,
    author = {Berry, Michael Victor},
    title = {Quantal phase factors accompanying adiabatic changes},
    journal = {Proceedings of the Royal Society of London. A. Mathematical and Physical Sciences},
    volume = {392},
    number = {1802},
    pages = {45-57},
    year = {1984},
    month = {03},
    issn = {0080-4630},
    doi = {10.1098/rspa.1984.0023},
    url = {https://doi.org/10.1098/rspa.1984.0023},
}

@article{gagliano1988dynamic,
  title = {Dynamic correlation functions in quantum many-body systems at zero temperature},
  author = {Gagliano, E. R. and Balseiro, C. A.},
  journal = {Phys. Rev. B},
  volume = {38},
  issue = {16},
  pages = {11766--11773},
  numpages = {0},
  year = {1988},
  month = {Dec},
  publisher = {American Physical Society},
  doi = {10.1103/PhysRevB.38.11766},
  url = {https://link.aps.org/doi/10.1103/PhysRevB.38.11766}
}

@article{white1992density,
  title = {Density matrix formulation for quantum renormalization groups},
  author = {White, Steven R.},
  journal = {Phys. Rev. Lett.},
  volume = {69},
  issue = {19},
  pages = {2863--2866},
  numpages = {0},
  year = {1992},
  month = {Nov},
  publisher = {American Physical Society},
  doi = {10.1103/PhysRevLett.69.2863},
  url = {https://link.aps.org/doi/10.1103/PhysRevLett.69.2863}
}

@article{dagotto1994correlated,
  title = {Correlated electrons in high-temperature superconductors},
  author = {Dagotto, Elbio},
  journal = {Rev. Mod. Phys.},
  volume = {66},
  issue = {3},
  pages = {763--840},
  numpages = {0},
  year = {1994},
  month = {Jul},
  publisher = {American Physical Society},
  doi = {10.1103/RevModPhys.66.763},
  url = {https://link.aps.org/doi/10.1103/RevModPhys.66.763}
}

@article{zhang1997constrained,
  title = {Constrained path {M}onte {C}arlo method for fermion ground states},
  author = {Zhang, Shiwei and Carlson, J. and Gubernatis, J. E.},
  journal = {Phys. Rev. B},
  volume = {55},
  issue = {12},
  pages = {7464--7477},
  numpages = {0},
  year = {1997},
  month = {Mar},
  publisher = {American Physical Society},
  doi = {10.1103/PhysRevB.55.7464},
  url = {https://link.aps.org/doi/10.1103/PhysRevB.55.7464}
}

@article{dzyaloshinskii2003some,
  title = {Some consequences of the {L}uttinger theorem: {T}he Luttinger surfaces in non-Fermi liquids and Mott insulators},
  author = {Dzyaloshinskii, Igor},
  journal = {Phys. Rev. B},
  volume = {68},
  issue = {8},
  pages = {085113},
  numpages = {6},
  year = {2003},
  month = {Aug},
  publisher = {American Physical Society},
  doi = {10.1103/PhysRevB.68.085113},
  url = {https://link.aps.org/doi/10.1103/PhysRevB.68.085113}
}

@article{zhang2003quantum,
  title = {Quantum {M}onte {C}arlo {M}ethod using {P}hase-{F}ree {R}andom {W}alks with {S}later {D}eterminants},
  author = {Zhang, Shiwei and Krakauer, Henry},
  journal = {Phys. Rev. Lett.},
  volume = {90},
  issue = {13},
  pages = {136401},
  numpages = {4},
  year = {2003},
  month = {Apr},
  publisher = {American Physical Society},
  doi = {10.1103/PhysRevLett.90.136401},
  url = {https://link.aps.org/doi/10.1103/PhysRevLett.90.136401}
}

@article{mancini2004hubbard,
    author = {F. Mancini and A. Avella},
    title = {The {H}ubbard model within the equations of motion approach},
    journal = {Advances in Physics},
    volume = {53},
    number = {5-6},
    pages = {537--768},
    year = {2004},
    publisher = {Taylor \& Francis},
    doi = {10.1080/00018730412331303722},
    URL = {https://doi.org/10.1080/00018730412331303722}
}

@article{schollwock2005density,
  title = {The density-matrix renormalization group},
  author = {Schollw\"ock, U.},
  journal = {Rev. Mod. Phys.},
  volume = {77},
  issue = {1},
  pages = {259--315},
  numpages = {0},
  year = {2005},
  month = {Apr},
  publisher = {American Physical Society},
  doi = {10.1103/RevModPhys.77.259},
  url = {https://link.aps.org/doi/10.1103/RevModPhys.77.259}
}

@article{SCHOLLWOCK201196,
    title = {The density-matrix renormalization group in the age of matrix product states},
    journal = {Annals of Physics},
    volume = {326},
    number = {1},
    pages = {96-192},
    year = {2011},
    issn = {0003-4916},
    doi = {https://doi.org/10.1016/j.aop.2010.09.012},
    url = {https://www.sciencedirect.com/science/article/pii/S0003491610001752},
    author = {Ulrich Schollwöck},
}

@article{stanescu2006fermi,
  title = {Fermi arcs and hidden zeros of the {G}reen function in the pseudogap state},
  author = {Stanescu, Tudor D. and Kotliar, Gabriel},
  journal = {Phys. Rev. B},
  volume = {74},
  issue = {12},
  pages = {125110},
  numpages = {6},
  year = {2006},
  month = {Sep},
  publisher = {American Physical Society},
  doi = {10.1103/PhysRevB.74.125110},
  url = {https://link.aps.org/doi/10.1103/PhysRevB.74.125110}
}

@article{yang2006phenomenological,
  title = {Phenomenological theory of the pseudogap state},
  author = {Yang, Kai-Yu and Rice, T. M. and Zhang, Fu-Chun},
  journal = {Phys. Rev. B},
  volume = {73},
  issue = {17},
  pages = {174501},
  numpages = {10},
  year = {2006},
  month = {May},
  publisher = {American Physical Society},
  doi = {10.1103/PhysRevB.73.174501},
  url = {https://link.aps.org/doi/10.1103/PhysRevB.73.174501}
}

@article{stanescu2007theory,
  title = {Theory of the {L}uttinger surface in doped {M}ott insulators},
  author = {Stanescu, Tudor D. and Phillips, Philip and Choy, Ting-Pong},
  journal = {Phys. Rev. B},
  volume = {75},
  issue = {10},
  pages = {104503},
  numpages = {9},
  year = {2007},
  month = {Mar},
  publisher = {American Physical Society},
  doi = {10.1103/PhysRevB.75.104503},
  url = {https://link.aps.org/doi/10.1103/PhysRevB.75.104503}
}

@Article{rosch2007breakdown,
    author={Rosch, A.},
    title={Breakdown of {L}uttinger's theorem in two-orbital {M}ott insulators},
    journal={The European Physical Journal B},
    year={2007},
    month={Oct},
    day={01},
    volume={59},
    number={4},
    pages={495-502},
    issn={1434-6036},
    doi={10.1140/epjb/e2007-00312-3},
    url={https://doi.org/10.1140/epjb/e2007-00312-3}
}

@article{sakai2009evolution,
  title = {Evolution of {E}lectronic {S}tructure of {D}oped {M}ott {I}nsulators: {R}econstruction of {P}oles and {Z}eros of {G}reen's {F}unction},
  author = {Sakai, Shiro and Motome, Yukitoshi and Imada, Masatoshi},
  journal = {Phys. Rev. Lett.},
  volume = {102},
  issue = {5},
  pages = {056404},
  numpages = {4},
  year = {2009},
  month = {Feb},
  publisher = {American Physical Society},
  doi = {10.1103/PhysRevLett.102.056404},
  url = {https://link.aps.org/doi/10.1103/PhysRevLett.102.056404}
}

@article{fidkowski2010effects,
  title = {Effects of interactions on the topological classification of free fermion systems},
  author = {Fidkowski, Lukasz and Kitaev, Alexei},
  journal = {Phys. Rev. B},
  volume = {81},
  issue = {13},
  pages = {134509},
  numpages = {9},
  year = {2010},
  month = {Apr},
  publisher = {American Physical Society},
  doi = {10.1103/PhysRevB.81.134509},
  url = {https://link.aps.org/doi/10.1103/PhysRevB.81.134509}
}

@article{wang2010topological,
  title = {Topological {O}rder {P}arameters for {I}nteracting {T}opological {I}nsulators},
  author = {Wang, Zhong and Qi, Xiao-Liang and Zhang, Shou-Cheng},
  journal = {Phys. Rev. Lett.},
  volume = {105},
  issue = {25},
  pages = {256803},
  numpages = {4},
  year = {2010},
  month = {Dec},
  publisher = {American Physical Society},
  doi = {10.1103/PhysRevLett.105.256803},
  url = {https://link.aps.org/doi/10.1103/PhysRevLett.105.256803}
}

@article{sakai2010doped,
  title = {Doped high-{${T}_{c}$} cuprate superconductors elucidated in the light of zeros and poles of the electronic {G}reen's function},
  author = {Sakai, Shiro and Motome, Yukitoshi and Imada, Masatoshi},
  journal = {Phys. Rev. B},
  volume = {82},
  issue = {13},
  pages = {134505},
  numpages = {16},
  year = {2010},
  month = {Oct},
  publisher = {American Physical Society},
  doi = {10.1103/PhysRevB.82.134505},
  url = {https://link.aps.org/doi/10.1103/PhysRevB.82.134505}
}

@article{xiao2010berry,
  title = {Berry phase effects on electronic properties},
  author = {Xiao, Di and Chang, Ming-Che and Niu, Qian},
  journal = {Rev. Mod. Phys.},
  volume = {82},
  issue = {3},
  pages = {1959--2007},
  numpages = {0},
  year = {2010},
  month = {Jul},
  publisher = {American Physical Society},
  doi = {10.1103/RevModPhys.82.1959},
  url = {https://link.aps.org/doi/10.1103/RevModPhys.82.1959}
}

@article{chen2011classification,
  title = {Classification of gapped symmetric phases in one-dimensional spin systems},
  author = {Chen, Xie and Gu, Zheng-Cheng and Wen, Xiao-Gang},
  journal = {Phys. Rev. B},
  volume = {83},
  issue = {3},
  pages = {035107},
  numpages = {19},
  year = {2011},
  month = {Jan},
  publisher = {American Physical Society},
  doi = {10.1103/PhysRevB.83.035107},
  url = {https://link.aps.org/doi/10.1103/PhysRevB.83.035107}
}

@article{chen2011unified,
  title = {Unified formalism for calculating polarization, magnetization, and more in a periodic insulator},
  author = {Chen, Kuang-Ting and Lee, Patrick A.},
  journal = {Phys. Rev. B},
  volume = {84},
  issue = {20},
  pages = {205137},
  numpages = {14},
  year = {2011},
  month = {Nov},
  publisher = {American Physical Society},
  doi = {10.1103/PhysRevB.84.205137},
  url = {https://link.aps.org/doi/10.1103/PhysRevB.84.205137}
}

@article{gurarie2011single,
  title = {Single-particle Green's functions and interacting topological insulators},
  author = {Gurarie, V.},
  journal = {Phys. Rev. B},
  volume = {83},
  issue = {8},
  pages = {085426},
  numpages = {15},
  year = {2011},
  month = {Feb},
  publisher = {American Physical Society},
  doi = {10.1103/PhysRevB.83.085426},
  url = {https://link.aps.org/doi/10.1103/PhysRevB.83.085426}
}

@article{manmana2012topological,
  title = {Topological invariants and interacting one-dimensional fermionic systems},
  author = {Manmana, Salvatore R. and Essin, Andrew M. and Noack, Reinhard M. and Gurarie, Victor},
  journal = {Phys. Rev. B},
  volume = {86},
  issue = {20},
  pages = {205119},
  numpages = {12},
  year = {2012},
  month = {Nov},
  publisher = {American Physical Society},
  doi = {10.1103/PhysRevB.86.205119},
  url = {https://link.aps.org/doi/10.1103/PhysRevB.86.205119}
}

@article{pollmann2012symmetry,
  title = {Symmetry protection of topological phases in one-dimensional quantum spin systems},
  author = {Pollmann, Frank and Berg, Erez and Turner, Ari M. and Oshikawa, Masaki},
  journal = {Phys. Rev. B},
  volume = {85},
  issue = {7},
  pages = {075125},
  numpages = {9},
  year = {2012},
  month = {Feb},
  publisher = {American Physical Society},
  doi = {10.1103/PhysRevB.85.075125},
  url = {https://link.aps.org/doi/10.1103/PhysRevB.85.075125}
}

@article{wang2012simplified,
  title = {{Simplified Topological Invariants for Interacting Insulators}},
  author = {Wang, Zhong and Zhang, Shou-Cheng},
  journal = {Phys. Rev. X},
  volume = {2},
  issue = {3},
  pages = {031008},
  numpages = {6},
  year = {2012},
  month = {Aug},
  publisher = {American Physical Society},
  doi = {10.1103/PhysRevX.2.031008},
  url = {https://link.aps.org/doi/10.1103/PhysRevX.2.031008}
}

@article{dave2013absence,
  title = {Absence of {L}uttinger's {T}heorem due to {Z}eros in the {S}ingle-{P}article {G}reen {F}unction},
  author = {Dave, Kiaran B. and Phillips, Philip W. and Kane, Charles L.},
  journal = {Phys. Rev. Lett.},
  volume = {110},
  issue = {9},
  pages = {090403},
  numpages = {5},
  year = {2013},
  month = {Feb},
  publisher = {American Physical Society},
  doi = {10.1103/PhysRevLett.110.090403},
  url = {https://link.aps.org/doi/10.1103/PhysRevLett.110.090403}
}

@article{nourafkan2013electric,
  title = {Electric polarization in correlated insulators},
  author = {Nourafkan, R. and Kotliar, G.},
  journal = {Phys. Rev. B},
  volume = {88},
  issue = {15},
  pages = {155121},
  numpages = {8},
  year = {2013},
  month = {Oct},
  publisher = {American Physical Society},
  doi = {10.1103/PhysRevB.88.155121},
  url = {https://link.aps.org/doi/10.1103/PhysRevB.88.155121}
}

@article{alexandradinata2014wilson,
  title = {Wilson-loop characterization of inversion-symmetric topological insulators},
  author = {Alexandradinata, A. and Dai, Xi and Bernevig, B. Andrei},
  journal = {Phys. Rev. B},
  volume = {89},
  issue = {15},
  pages = {155114},
  numpages = {18},
  year = {2014},
  month = {Apr},
  publisher = {American Physical Society},
  doi = {10.1103/PhysRevB.89.155114},
  url = {https://link.aps.org/doi/10.1103/PhysRevB.89.155114}
}

@article{yoshida2014characterization,
  title = {Characterization of a {T}opological {M}ott Insulator in {O}ne {D}imension},
  author = {Yoshida, Tsuneya and Peters, Robert and Fujimoto, Satoshi and Kawakami, Norio},
  journal = {Phys. Rev. Lett.},
  volume = {112},
  issue = {19},
  pages = {196404},
  numpages = {5},
  year = {2014},
  month = {May},
  publisher = {American Physical Society},
  doi = {10.1103/PhysRevLett.112.196404},
  url = {https://link.aps.org/doi/10.1103/PhysRevLett.112.196404}
}

@article{senthil2015symmetry,
   author = "Senthil, T.",
   title = "Symmetry-{P}rotected {T}opological {P}hases of {Q}uantum {M}atter", 
   journal= "Annual Review of Condensed Matter Physics",
   year = "2015",
   volume = "6",
   number = "Volume 6, 2015",
   pages = "299-324",
   doi = "https://doi.org/10.1146/annurev-conmatphys-031214-014740",
   url = "https://www.annualreviews.org/content/journals/10.1146/annurev-conmatphys-031214-014740",
   publisher = "Annual Reviews",
   issn = "1947-5462",
   type = "Journal Article",
}

@article{chiu2016classification,
  title = {Classification of topological quantum matter with symmetries},
  author = {Chiu, Ching-Kai and Teo, Jeffrey C. Y. and Schnyder, Andreas P. and Ryu, Shinsei},
  journal = {Rev. Mod. Phys.},
  volume = {88},
  issue = {3},
  pages = {035005},
  numpages = {63},
  year = {2016},
  month = {Aug},
  publisher = {American Physical Society},
  doi = {10.1103/RevModPhys.88.035005},
  url = {https://link.aps.org/doi/10.1103/RevModPhys.88.035005}
}

@article{he2016topological1,
  title = {Topological invariants for interacting topological insulators. {I}. {E}fficient numerical evaluation scheme and implementations},
  author = {He, Yuan-Yao and Wu, Han-Qing and Meng, Zi Yang and Lu, Zhong-Yi},
  journal = {Phys. Rev. B},
  volume = {93},
  issue = {19},
  pages = {195163},
  numpages = {18},
  year = {2016},
  month = {May},
  publisher = {American Physical Society},
  doi = {10.1103/PhysRevB.93.195163},
  url = {https://link.aps.org/doi/10.1103/PhysRevB.93.195163}
}

@article{he2016topological2,
  title = {Topological invariants for interacting topological insulators. {II}. {B}reakdown of single-particle {G}reen's function formalism},
  author = {He, Yuan-Yao and Wu, Han-Qing and Meng, Zi Yang and Lu, Zhong-Yi},
  journal = {Phys. Rev. B},
  volume = {93},
  issue = {19},
  pages = {195164},
  numpages = {13},
  year = {2016},
  month = {May},
  publisher = {American Physical Society},
  doi = {10.1103/PhysRevB.93.195164},
  url = {https://link.aps.org/doi/10.1103/PhysRevB.93.195164}
}

@article{queiroz2016dimensional,
  title = {{Dimensional Hierarchy of Fermionic Interacting Topological Phases}},
  author = {Queiroz, Raquel and Khalaf, Eslam and Stern, Ady},
  journal = {Phys. Rev. Lett.},
  volume = {117},
  issue = {20},
  pages = {206405},
  numpages = {5},
  year = {2016},
  month = {Nov},
  publisher = {American Physical Society},
  doi = {10.1103/PhysRevLett.117.206405},
  url = {https://link.aps.org/doi/10.1103/PhysRevLett.117.206405}
}

@article{benalcazar2017quantized,
    author = {Wladimir A. Benalcazar  and B. Andrei Bernevig  and Taylor L. Hughes },
    title = {Quantized electric multipole insulators},
    journal = {Science},
    volume = {357},
    number = {6346},
    pages = {61-66},
    year = {2017},
    doi = {10.1126/science.aah6442},
    URL = {https://www.science.org/doi/abs/10.1126/science.aah6442},
}

@article{seki2017topological,
  title = {Topological interpretation of the {L}uttinger theorem},
  author = {Seki, Kazuhiro and Yunoki, Seiji},
  journal = {Phys. Rev. B},
  volume = {96},
  issue = {8},
  pages = {085124},
  numpages = {22},
  year = {2017},
  month = {Aug},
  publisher = {American Physical Society},
  doi = {10.1103/PhysRevB.96.085124},
  url = {https://link.aps.org/doi/10.1103/PhysRevB.96.085124}
}

@article{verresen2017one,
  title = {One-dimensional symmetry protected topological phases and their transitions},
  author = {Verresen, Ruben and Moessner, Roderich and Pollmann, Frank},
  journal = {Phys. Rev. B},
  volume = {96},
  issue = {16},
  pages = {165124},
  numpages = {23},
  year = {2017},
  month = {Oct},
  publisher = {American Physical Society},
  doi = {10.1103/PhysRevB.96.165124},
  url = {https://link.aps.org/doi/10.1103/PhysRevB.96.165124}
}

@article{scheurer2018topological,
    author = {Mathias S. Scheurer  and Shubhayu Chatterjee  and Wei Wu  and Michel Ferrero  and Antoine Georges  and Subir Sachdev },
    title = {Topological order in the pseudogap metal},
    journal = {Proceedings of the National Academy of Sciences},
    volume = {115},
    number = {16},
    pages = {E3665-E3672},
    year = {2018},
    doi = {10.1073/pnas.1720580115},
    URL = {https://www.pnas.org/doi/abs/10.1073/pnas.1720580115},
}

@article{else2020topological,
  title = {Topological theory of {L}ieb-{S}chultz-{M}attis theorems in quantum spin systems},
  author = {Else, Dominic V. and Thorngren, Ryan},
  journal = {Phys. Rev. B},
  volume = {101},
  issue = {22},
  pages = {224437},
  numpages = {35},
  year = {2020},
  month = {Jun},
  publisher = {American Physical Society},
  doi = {10.1103/PhysRevB.101.224437},
  url = {https://link.aps.org/doi/10.1103/PhysRevB.101.224437}
}

@article{iraola2021towards,
  title = {Towards a topological quantum chemistry description of correlated systems: {T}he case of the {H}ubbard diamond chain},
  author = {Iraola, Mikel and Heinsdorf, Niclas and Tiwari, Apoorv and Lessnich, Dominik and Mertz, Thomas and Ferrari, Francesco and Fischer, Mark H. and Winter, Stephen M. and Pollmann, Frank and Neupert, Titus and Valent\'{\i}, Roser and Vergniory, Maia G.},
  journal = {Phys. Rev. B},
  volume = {104},
  issue = {19},
  pages = {195125},
  numpages = {17},
  year = {2021},
  month = {Nov},
  publisher = {American Physical Society},
  doi = {10.1103/PhysRevB.104.195125},
  url = {https://link.aps.org/doi/10.1103/PhysRevB.104.195125}
}

@article{lessnich2021elementary,
  title = {Elementary band representations for the single-particle {G}reen's function of interacting topological insulators},
  author = {Lessnich, Dominik and Winter, Stephen M. and Iraola, Mikel and Vergniory, Maia G. and Valent\'{\i}, Roser},
  journal = {Phys. Rev. B},
  volume = {104},
  issue = {8},
  pages = {085116},
  numpages = {11},
  year = {2021},
  month = {Aug},
  publisher = {American Physical Society},
  doi = {10.1103/PhysRevB.104.085116},
  url = {https://link.aps.org/doi/10.1103/PhysRevB.104.085116}
}

@article{fuchs2021orbital,
  title = {Orbital embedding and topology of one-dimensional two-band insulators},
  author = {Fuchs, Jean-No\"el and Pi\'echon, Fr\'ed\'eric},
  journal = {Phys. Rev. B},
  volume = {104},
  issue = {23},
  pages = {235428},
  numpages = {9},
  year = {2021},
  month = {Dec},
  publisher = {American Physical Society},
  doi = {10.1103/PhysRevB.104.235428},
  url = {https://link.aps.org/doi/10.1103/PhysRevB.104.235428}
}

@article{itensor,
	title={{The ITensor Software Library for Tensor Network Calculations}},
	author={Matthew Fishman and Steven R. White and E. Miles Stoudenmire},
	journal={SciPost Phys. Codebases},
	pages={4},
	year={2022},
	publisher={SciPost},
	doi={10.21468/SciPostPhysCodeb.4},
	url={https://scipost.org/10.21468/SciPostPhysCodeb.4}
}

@Article{fabrizio2022emergent,
    author={Fabrizio, Michele},
    title={Emergent quasiparticles at {L}uttinger surfaces},
    journal={Nature Communications},
    year={2022},
    month={Mar},
    day={23},
    volume={13},
    number={1},
    pages={1561},
    issn={2041-1723},
    doi={10.1038/s41467-022-29190-y},
    url={https://doi.org/10.1038/s41467-022-29190-y}
}

@Article{guzman2022geometry,
	title={{Geometry and topology tango in ordered and amorphous chiral matter}},
	author={Marcelo Guzmán and Denis Bartolo and David Carpentier},
	journal={SciPost Phys.},
	volume={12},
	pages={038},
	year={2022},
	publisher={SciPost},
	doi={10.21468/SciPostPhys.12.1.038},
	url={https://scipost.org/10.21468/SciPostPhys.12.1.038},
}

@article{skolimowski2022luttinger,
  title = {Luttinger's theorem in the presence of {L}uttinger surfaces},
  author = {Skolimowski, Jan and Fabrizio, Michele},
  journal = {Phys. Rev. B},
  volume = {106},
  issue = {4},
  pages = {045109},
  numpages = {13},
  year = {2022},
  month = {Jul},
  publisher = {American Physical Society},
  doi = {10.1103/PhysRevB.106.045109},
  url = {https://link.aps.org/doi/10.1103/PhysRevB.106.045109}
}

@article{blason2023unified,
  title = {Unified role of {G}reen's function poles and zeros in correlated topological insulators},
  author = {Blason, Andrea and Fabrizio, Michele},
  journal = {Phys. Rev. B},
  volume = {108},
  issue = {12},
  pages = {125115},
  numpages = {10},
  year = {2023},
  month = {Sep},
  publisher = {American Physical Society},
  doi = {10.1103/PhysRevB.108.125115},
  url = {https://link.aps.org/doi/10.1103/PhysRevB.108.125115}
}

@article{fabrizio2023spin,
  title = {Spin-{L}iquid {I}nsulators {C}an {B}e {L}andau's {F}ermi {L}iquids},
  author = {Fabrizio, Michele},
  journal = {Phys. Rev. Lett.},
  volume = {130},
  issue = {15},
  pages = {156702},
  numpages = {6},
  year = {2023},
  month = {Apr},
  publisher = {American Physical Society},
  doi = {10.1103/PhysRevLett.130.156702},
  url = {https://link.aps.org/doi/10.1103/PhysRevLett.130.156702}
}

@Article{jiang2023mottness,
	title={{Geometry and topology tango in ordered and amorphous chiral matter}},
	author={Marcelo Guzmán and Denis Bartolo and David Carpentier},
	journal={SciPost Phys.},
	volume={12},
	pages={038},
	year={2022},
	publisher={SciPost},
	doi={10.21468/SciPostPhys.12.1.038},
	url={https://scipost.org/10.21468/SciPostPhys.12.1.038},
}

@article{peralta2023connecting,
  title = {Connecting the Many-Body Chern Number to Luttinger's Theorem through St\ifmmode \check{r}\else \v{r}\fi{}eda's Formula},
  author = {Peralta Gavensky, Lucila and Sachdev, Subir and Goldman, Nathan},
  journal = {Phys. Rev. Lett.},
  volume = {131},
  issue = {23},
  pages = {236601},
  numpages = {7},
  year = {2023},
  month = {Dec},
  publisher = {American Physical Society},
  doi = {10.1103/PhysRevLett.131.236601},
  url = {https://link.aps.org/doi/10.1103/PhysRevLett.131.236601}
}

@Article{wagner2023mott,
    author={Wagner, N. and Crippa, L. and Amaricci, A. and Hansmann, P. and Klett, M. and K{\"o}nig, E. J. and Sch{\"a}fer, T. and Sante, D. Di and Cano, J. and Millis, A. J. and Georges, A. and Sangiovanni, G.},
    title={Mott insulators with boundary zeros},
    journal={Nature Communications},
    year={2023},
    month={Nov},
    day={20},
    volume={14},
    number={1},
    pages={7531},
    issn={2041-1723},
    doi={10.1038/s41467-023-42773-7},
    url={https://doi.org/10.1038/s41467-023-42773-7}
}

@article{zhao2023failure,
  title = {Failure of {T}opological {I}nvariants in {S}trongly {C}orrelated {M}atter},
  author = {Zhao, Jinchao and Mai, Peizhi and Bradlyn, Barry and Phillips, Philip},
  journal = {Phys. Rev. Lett.},
  volume = {131},
  issue = {10},
  pages = {106601},
  numpages = {7},
  year = {2023},
  month = {Sep},
  publisher = {American Physical Society},
  doi = {10.1103/PhysRevLett.131.106601},
  url = {https://link.aps.org/doi/10.1103/PhysRevLett.131.106601}
}

@article{blason2024luttinger,
  title = {Luttinger surface dominance and {F}ermi liquid behavior of topological {K}ondo insulators {${\mathrm{SmB}}_{6}$ and ${\mathrm{YbB}}_{12}$}},
  author = {Blason, Andrea and Pasqua, Ivan and Ferrero, Michel and Fabrizio, Michele},
  journal = {Phys. Rev. B},
  volume = {110},
  issue = {23},
  pages = {235115},
  numpages = {13},
  year = {2024},
  month = {Dec},
  publisher = {American Physical Society},
  doi = {10.1103/PhysRevB.110.235115},
  url = {https://link.aps.org/doi/10.1103/PhysRevB.110.235115}
}

@article{bollmann2024topological,
  title = {Topological {G}reen's {F}unction {Z}eros in an {E}xactly {S}olved {M}odel and {B}eyond},
  author = {Bollmann, Steffen and Setty, Chandan and Seifert, Urban F. P. and K\"onig, Elio J.},
  journal = {Phys. Rev. Lett.},
  volume = {133},
  issue = {13},
  pages = {136504},
  numpages = {7},
  year = {2024},
  month = {Sep},
  publisher = {American Physical Society},
  doi = {10.1103/PhysRevLett.133.136504},
  url = {https://link.aps.org/doi/10.1103/PhysRevLett.133.136504}
}

@article{setty2024symmetry,
  title = {Symmetry constraints and spectral crossing in a {M}ott insulator with {G}reen's function zeros},
  author = {Setty, Chandan and Sur, Shouvik and Chen, Lei and Xie, Fang and Hu, Haoyu and Paschen, Silke and Cano, Jennifer and Si, Qimiao},
  journal = {Phys. Rev. Res.},
  volume = {6},
  issue = {3},
  pages = {L032018},
  numpages = {6},
  year = {2024},
  month = {Jul},
  publisher = {American Physical Society},
  doi = {10.1103/PhysRevResearch.6.L032018},
  url = {https://link.aps.org/doi/10.1103/PhysRevResearch.6.L032018}
}

@article{setty2024electronic,
  title = {Electronic properties, correlated topology, and {G}reen's function zeros},
  author = {Setty, Chandan and Xie, Fang and Sur, Shouvik and Chen, Lei and Vergniory, Maia G. and Si, Qimiao},
  journal = {Phys. Rev. Res.},
  volume = {6},
  issue = {3},
  pages = {033235},
  numpages = {15},
  year = {2024},
  month = {Sep},
  publisher = {American Physical Society},
  doi = {10.1103/PhysRevResearch.6.033235},
  url = {https://link.aps.org/doi/10.1103/PhysRevResearch.6.033235}
}

@article{wagner2024edge,
  title = {Edge {Z}eros and {B}oundary {S}pinons in {T}opological {M}ott {I}nsulators},
  author = {Wagner, Niklas and Guerci, Daniele and Millis, Andrew J. and Sangiovanni, Giorgio},
  journal = {Phys. Rev. Lett.},
  volume = {133},
  issue = {12},
  pages = {126504},
  numpages = {7},
  year = {2024},
  month = {Sep},
  publisher = {American Physical Society},
  doi = {10.1103/PhysRevLett.133.126504},
  url = {https://link.aps.org/doi/10.1103/PhysRevLett.133.126504}
}

@article{worm2024fermi,
  title = {Fermi and {L}uttinger {A}rcs: {T}wo {C}oncepts, {R}ealized on {O}ne {S}urface},
  author = {Worm, Paul and Reitner, Matthias and Held, Karsten and Toschi, Alessandro},
  journal = {Phys. Rev. Lett.},
  volume = {133},
  issue = {16},
  pages = {166501},
  numpages = {7},
  year = {2024},
  month = {Oct},
  publisher = {American Physical Society},
  doi = {10.1103/PhysRevLett.133.166501},
  url = {https://link.aps.org/doi/10.1103/PhysRevLett.133.166501}
}

@misc{aligia2025probing,
      title={Probing {B}oundary {S}pins in the {S}u-{S}chrieffer-{H}eeger-{H}ubbard model}, 
      author={Armando A. Aligia and Alejandro M. Lobos and Lucila Peralta Gavensky and Claudio J. Gazza},
      year={2025},
      eprint={2511.17173},
      archivePrefix={arXiv},
      primaryClass={cond-mat.str-el},
      url={https://arxiv.org/abs/2511.17173}, 
}

@misc{dionne2025characterizing,
      title={Characterizing {M}ott {I}nsulators in the {I}nteracting {O}ne-{B}ody {P}icture}, 
      author={Theo N. Dionne and Santiago Villodre and Mikel Iraola and Maia G. Vergniory},
      year={2026},
      eprint={2511.07331},
      archivePrefix={arXiv},
      primaryClass={cond-mat.str-el},
      url={https://arxiv.org/abs/2511.07331}, 
}

@article{haurie2024bands,
    doi = {10.1088/1361-648X/ad1e07},
    url = {https://doi.org/10.1088/1361-648X/ad1e07},
    year = {2024},
    month = {mar},
    publisher = {IOP Publishing},
    volume = {36},
    number = {25},
    pages = {255601},
    author = {Haurie, L and Grandadam, M and Pangburn, E and Banerjee, A and Burdin, S and Pépin, C},
    title = {Bands renormalization and superconductivity in the strongly correlated {H}ubbard model using composite operators method},
    journal = {Journal of Physics: Condensed Matter},
}

@article{banerjee2025charge,
  title = {Charge density wave solutions of the {H}ubbard model in the composite operator formalism},
  author = {Banerjee, Anurag and Pangburn, Emile and Mahato, Chiranjit and Ghosal, Amit and P\'epin, Catherine},
  journal = {Phys. Rev. B},
  volume = {111},
  issue = {16},
  pages = {165123},
  numpages = {15},
  year = {2025},
  month = {Apr},
  publisher = {American Physical Society},
  doi = {10.1103/PhysRevB.111.165123},
  url = {https://link.aps.org/doi/10.1103/PhysRevB.111.165123}
}

@Article{pasqua2025fermi,
	title={{Fermi-liquid corrections to the intrinsic anomalous {H}all conductivity of topological metals}},
	author={Ivan Pasqua and Michele Fabrizio},
	journal={SciPost Phys.},
	volume={19},
	pages={014},
	year={2025},
	publisher={SciPost},
	doi={10.21468/SciPostPhys.19.1.014},
	url={https://scipost.org/10.21468/SciPostPhys.19.1.014},
}

@article{pangburn2025topological,
  title = {Topological charge excitations and {G}reen's function zeros in paramagnetic {M}ott insulators},
  author = {Pangburn, Emile and P\'epin, Catherine and Banerjee, Anurag},
  journal = {Phys. Rev. B},
  volume = {112},
  issue = {8},
  pages = {085105},
  numpages = {18},
  year = {2025},
  month = {Aug},
  publisher = {American Physical Society},
  doi = {10.1103/4k4y-4hj4},
  url = {https://link.aps.org/doi/10.1103/4k4y-4hj4}
}

@article{lehmann2025probing,
  title = {Probing {G}reen's {F}unction {Z}eros by {C}otunneling through {M}ott {I}nsulators},
  author = {Lehmann, Carl and Crippa, Lorenzo and Sangiovanni, Giorgio and Budich, Jan Carl},
  journal = {Phys. Rev. Lett.},
  volume = {135},
  issue = {10},
  pages = {106303},
  numpages = {8},
  year = {2025},
  month = {Sep},
  publisher = {American Physical Society},
  doi = {10.1103/jnq4-sykq},
  url = {https://link.aps.org/doi/10.1103/jnq4-sykq}
}

@misc{konig2026topological,
      title={{Topological Floquet Green's function zeros}}, 
      author={Elio J. König and Aditi Mitra},
      year={2026},
      eprint={2602.21199},
      archivePrefix={arXiv},
      primaryClass={cond-mat.mes-hall},
      url={https://arxiv.org/abs/2602.21199}, 
}

@misc{markov2026luttinger,
      title={{Luttinger's Theorem Violation and Green's Function Topological Invariants in a Fractional Chern Insulator}}, 
      author={Anton A. Markov and Andrey M. Nikishin and Nigel R. Cooper and Nathan Goldman and Lucila Peralta Gavensky},
      year={2026},
      eprint={2603.17006},
      archivePrefix={arXiv},
      primaryClass={cond-mat.str-el},
      url={https://arxiv.org/abs/2603.17006}, 
}

@unpublished{safire26,
    author = {{B. Eskridge \textit{et al.}}},
    title = {{private communication; SAFIRE}},
    year = {2026},
    note = {\href{https://safire.flatironinstitute.org/}{https://safire.flatironinstitute.org/}}
}

@article{kogel2026poles,
  title={Poles-zeros duality in semi-holographic Mott insulators},
  author={K{\"o}gel, Thomas and Caddeo, Alessio and Pitters, Amelie and Paoletti, Francesca and Crippa, Lorenzo and Sangiovanni, Giorgio and Meyer, Ren{\'e} and Erdmenger, Johanna},
  journal={arXiv preprint arXiv:2605.20321},
  year={2026}
}

@article{mitra2026signatures,
  title={Signatures of Green's function zeros and their topology using impurity spectroscopy},
  author={Mitra, Sayan and Xie, Fang and Kolmer, Marek and Si, Qimiao and Setty, Chandan},
  journal={arXiv preprint arXiv:2602.23477},
  year={2026}
}

\appendix

\pagebreak
~
\newpage

\setcounter{equation}{0}
\setcounter{figure}{0}
\setcounter{table}{0}
\setcounter{page}{1}

\renewcommand{\thefigure}{S\arabic{figure}}
\renewcommand{\bibnumfmt}[1]{[S#1]}
\renewcommand{\citenumfont}[1]{S#1}

\onecolumngrid
\begin{center}

{\large\textbf{\boldmath
Supplemental Materials\\ [0.5em] {\small to} \\ [0.5em]
Fundamental limitations of single-particle Green's-function zeroes as probes of many-body topology}}\\[1.5em]

Emile~\surname{Pangburn}$^{1,2}$, Olivier~\surname{Gingras\,\orcidlink{0000-0003-3970-6273}}$^{2,1}$\\[0.5em]

\textit{\small
$^1$Institut de Physique Théorique, Université Paris Saclay, CEA CNRS, Orme des Merisiers, 91190 Gif-sur-Yvette Cedex, France\\
$^2$Center for Computational Quantum Physics, Flatiron Institute, 162 5th Avenue, New York, New York 10010, USA
}
\vspace{2em}
\end{center}

\twocolumngrid

\setcounter{secnumdepth}{3}

\section{Review of Green's function's topological invariants}
\label{Sec:Review}

We investigate the correspondence between single-particle Green’s-function (GF) topology and ground-state topology in one dimension. We focus on bulk-gapped systems at half filling, which become Mott insulators in the strong-interaction regime. Because of the absence of intrinsic topological order in one dimension, the topological classification of these models is determined by the nontrivial projective representations carried by the edge states, as classified by group cohomology~\cite{chen2011unified, verresen2017one, else2020topological}. In more intuitive terms, this classification is encoded in how the edge states transform under the action of symmetries. For example, a spin-singlet transforms trivially under $SO(3)$ rotation symmetry, whereas a spin-doublet transforms non-trivially.

To probe the topology of these states, we compute the many-body excitation gap under both periodic and open boundary conditions (PBC and OBC). A finite many-body gap with PBC, accompanied by a vanishing many-body gap with OBC, signals a topological phase. In the following, we restrict our analysis to translation-invariant systems with $SO(3)$ rotation symmetry.

This well-established understanding of many-body ground-state topology in one dimension will then be compared with the topology inferred from single-particle GF invariants.

\subsection{Lehmann representation of the Green's function}

We study the relationship between the topologies of the ground state and of the GF in fermionic models. In particular, we focus on single-particle retarded ($R$) GF defined as
\begin{align}
    \mathcal{G}^R_{ij}(t) = -i \theta(t)\Big\langle\left\{c^\dagger_{j}(t),c_{i}\right\}\Big\rangle,
\end{align}
where $c_{i}$ ($c^\dagger_{i}$) is the annihilation (creation) operator of an electron with quantum numbers $i$ which may include lattice site, orbital, layer, or spin indices, $A(t) \equiv e^{iHt}Ae^{-iHt}$, $\theta$ is the Heaviside function, $\{\cdot,\cdot\}$ is the anticommutator and $\langle\cdot,\cdot \rangle$ is the thermal expectation value. In the following, we consider only retarded GF, and the superscript $R$ is omitted. For a system with $N$ electrons, the retarded GF admits a Lehmann representation:
\begin{align}
    \nonumber \mathcal{G}_{ij}(\omega) & = \sum\limits_n \dfrac{\langle \Psi^N_0|c_i|n^{N+1} \rangle \langle n^{N+1}|c_j^\dagger|\Psi^N_0\rangle}{\omega-(E_n^{N+1}-E_0^N)+i0^+}\\
    & + \sum\limits_m \dfrac{\langle \Psi^N_0|c_j^\dagger| m^{N-1}\rangle \langle m^{N-1}|c_i|\Psi^N_0\rangle}{\omega+(E_m^{N-1}-E_0^N)+i0^+},
    \label{eq:lehmann}
\end{align}
where $|\Psi^N_0\rangle$ is the $N$-particle ground state with energy $E^N_0$, and $|n^{N+1}\rangle$ ($|m^{N-1}\rangle$) denote excited states with one additional (one fewer) electron and with energy $E^{N+1}_n$ ($E^{N-1}_m$).

The poles of the GF correspond to the addition or removal of an electron and therefore describe quasiparticle excitations. These poles determine the coherent part of the single-particle excitation spectrum. In contrast, zeroes of the GF occur when the GF possess a zero mode. Such zeroes arise naturally in interacting systems, particularly in Hubbard-type models, where the electron fractionalizes into distinct charge excitations (holons and doublons).

\subsection{Existence of Green's function zeroes\label{App:Greenzeroes}}
We now discuss the relationship between the number of poles and zeroes of the single-particle GF within a fixed set of quantum numbers. Our discussion follows Ref.~\citenum{seki2017topological}.

For diagonal GF elements associated with conserved quantum numbers, the difference between the number of poles and zeroes is invariant under continuous changes of interaction strength~\cite{seki2017topological}.  This follows from the monotonic behavior of the real part of the GF between its poles, and from its divergence and sign change at each pole. As interactions are introduced and additional many-body excitations emerge, this constraint implies the appearance of GF zeroes. A detailed demonstration of these properties is provided now. We focus on the diagonal elements without loss of generality, as the single-particle GF can always be diagonalized to reduce the problem to this simplified setting.

The real part of the GF is a strictly monotonic function of $\omega$ away from the poles. Indeed, differentiating the Lehmann representation Eq.~\ref{eq:lehmann} yields, for the GF diagonal elements,
\begin{align}
\partial_\omega \mathcal{G}_{ii}(\omega)
=
&-\sum_n
\dfrac{|\langle \Psi_0 | c_i | n^{N+1} \rangle|^2}
{\left[\omega-(E_n^{N+1}-E_0^N)+i\eta\right]^2}
\nonumber &\\
&-
\sum_m
\dfrac{|\langle \Psi_0 | c_i^\dagger | m^{N-1} \rangle|^2}
{\left[\omega+(E_m^{N-1}-E_0^N)+i\eta\right]^2}.
\end{align}

\begin{figure}[t!]
    \centering
    \includegraphics[width=1.0\linewidth]{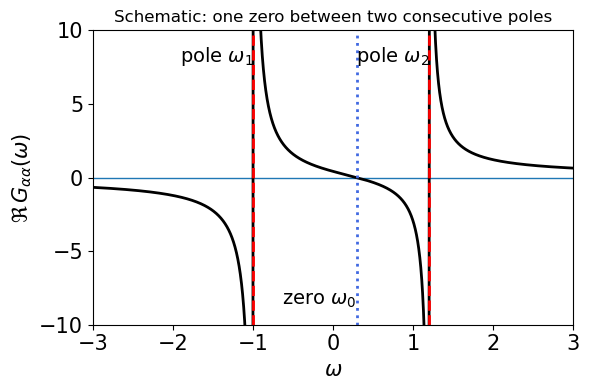}
    \caption{
        \textbf{Schematic illustration of the real part of the retarded Green’s function.}
        $\Re\mathcal{G}_{\alpha\alpha}(\omega)$ between two consecutive poles at $\omega_1$. Due to its monotonic behavior between poles, $\Re\mathcal{G}_{\alpha\alpha}(\omega)$ necessarily crosses zero once at $\omega_0$. The black curve shows $\Re \mathcal{G}_{\alpha\alpha}$, while red and blue dotted vertical lines mark poles and zeroes, respectively.}
    \label{fig:existence_zeroes}
\end{figure}

Since each term contributes a strictly negative real part, we conclude that 
for all $\omega$ away from the poles, the diagonal elements real part are monotonically decreasing. Now consider two consecutive poles at $\omega_1$ and $\omega_2$, with $\omega_1<\omega_2$. To the right of the first pole one has $\Re\mathcal{G}_{\alpha\alpha}(\omega_1^{+})>0$, while to the left of the second pole $\Re\mathcal{G}_{\alpha\alpha}(\omega_2^{-})<0$.
Because $\Re \mathcal{G}_{\alpha\alpha}(\omega)$ is continuous and strictly monotonic between poles, it must vanish exactly once in the interval $]\omega_1,\omega_2[$. Hence, between any two consecutive poles of the diagonal GF there exists precisely one zero. This establishes that, for each diagonal component $\mathcal{G}_{\alpha\alpha}$, the difference between the number of poles and the number of zeroes is a conserved quantity.

\subsection{Determinant of the Green's function}

In practice, GF zeroes are identified by examining the determinant of the single-particle GF. A zero occurs at a frequency $\omega$ when this determinant vanishes, indicating that the GF possesses a zero mode at that energy. Say $w_z$ is a zero, then
\begin{align}
    |\det\mathcal{G}(\omega_z)|=0.
\end{align}

GF zeroes can also be understood as divergences of the self-energy $\Sigma(\omega)$. Expressing $\mathcal{G}$ with the Dyson equation, we have
\begin{align}
    \mathcal{G}^R(\mathbf{k},\omega) = \left(\omega+i0^+-H_0(\mathbf{k})-\Sigma(\mathbf{k},\omega)\right)^{-1},
\end{align}
where $H_0(\mathbf{k})$ is the non-interacting part of the system's Hamiltonian. It is straightforward to see that zeroes correspond to divergence of the self-energy.

\subsection{Chiral symmetry and topological invariants}
The topology of the SSH chain is protected by sublattice-symmetry, allowing the definition of topological invariants. This symmetry acts on the many-body Hamiltonian $H$ as 
\begin{align}
    \label{eq:chiral_sym}
    \hat{\mathcal{S}} H \hat{\mathcal{S}}^{-1} = H, \quad \text{where} \quad 
    \hat{\mathcal{S}}^{-1} c_{i\sigma}\, \hat{\mathcal{S}} = (-1)^i c^\dagger_{i\sigma}.
\end{align}
This transformation assigns opposite signs to the two sublattices, reflecting the bipartite structure of the model. The non-interacting part of the Hamiltonian can be written in the sublatti98
\begin{align}
    H_0 =\sum\limits_{\textbf{k}\alpha\beta} \left[h_0(\textbf{k})\right]_{\alpha\beta} c^\dagger_{\textbf{k}\alpha}c_{\textbf{k}\beta}&
\end{align}

with $\textbf{k}$ the conserved momentum and $\alpha,\beta$ sublattices index. $h_0(\textbf{k})$ is the Bloch Hamiltonian. Acting with $\hat{S}$ on $H_0$ as in Eq.~(\ref{eq:chiral_sym}) leads to the following identity:
\begin{align}
    H_0 = -\sum\limits_{\textbf{k}\alpha\beta}\left[\sigma_z\right]_{\alpha\gamma}\left[h_0(\textbf{k})\right]_{\gamma\delta}\left[\sigma_z\right]_{\delta\beta}c^\dagger_{\textbf{k}\alpha}c_{\textbf{k}\beta}&
\end{align}
where repeated indices are summed, $\sigma_z$ is the third Pauli matrix.
Consequently, this imposes the following constraint on the Bloch Hamiltonian:
\begin{align}
    h_0(\textbf{k}) = -\sigma_z h_0(\textbf{k})\sigma_z .
\end{align}
Under this constraint, $h_0(\textbf{k})$ can be continuously deformed and spectrally flattened into a purely off-diagonal form,
\begin{align}
    h_0(\textbf{k}) =
    \begin{pmatrix}
        0 & Q(\textbf{k}) \\
        Q^\dagger(\textbf{k}) & 0 \\
    \end{pmatrix},
\end{align}
where $Q(\textbf{k})$ is a unitary matrix. The topological class of $Q(\textbf{k})$ in the absence of other symmetries is the winding number, that is
\begin{align}
    \nu = \frac{1}{2\pi} \int d\textbf{k} \ \mathrm{Tr}\left[ Q(\textbf{k}) \partial_\textbf{k} Q(\textbf{k}) \right].
    \label{eq:NI_nu}
\end{align}
This invariant classifies one-dimensional chiral-symmetric insulators~\cite{chiu2016classification}. As a technical remark, to be well-defined, the winding number has to be computed with an embedding in which all orbitals are placed at the same position within the unit cell~\cite{fuchs2021orbital, guzman2022geometry}.

\section{Minimal examples: Hubbard dot and dimer\label{Sec:Dot_Dimer}}

To understand intuitively the position of GF zeroes in frequency space, it is useful to start with the simplest interacting systems: a single Hubbard dot and a two-site Hubbard dimer. These models can be solved exactly, providing insight into how spectral-weight splitting produces GF zeroes. Moreover, these two examples are relevant to understand the zeroes boundary spectrum in lattice models.

\subsection{Single site: Hubbard dot}
The Hubbard Hamiltonian on a single site $1$ is given by 
\begin{align}
    \label{eq:ham_dot}
    H_{\text{dot}} = U\left(n_{1\uparrow}-\tfrac12\right)\left(n_{1\downarrow}-\tfrac12\right)
\end{align}
with $n_{1\sigma}\equiv c^\dagger_{1\sigma}c_{1\sigma}$. Its Hilbert space consists in the following states: $\{|0\rangle, |\uparrow\rangle, |\downarrow\rangle, |\uparrow\downarrow\rangle\}$.

The electronic operator on a single interacting site can be decomposed exactly as
\begin{align}
c_{1\sigma}=\xi_{1\sigma}+\eta_{1\sigma},
\end{align}
where 
\begin{align}
\xi_{1\sigma} = c_{1\sigma}\left(1-n_{1\overline{\sigma}}\right) \quad \text{and} \quad 
\eta_{i\sigma} = c_{i\sigma}n_{i\overline{\sigma}},
\end{align}
with $\xi_{1\sigma}$ ($\xi_{1\sigma}^\dagger$) the annihilation (creation) operator of an electron on a singly occupied (empty) site, and 
$\eta_{i\sigma}$ ($\eta^\dagger_{i\sigma}$) the annihilation (creation) operator of an electron on a doubly (singly) occupied site. 

\begin{figure}
    \includegraphics[width=\linewidth]{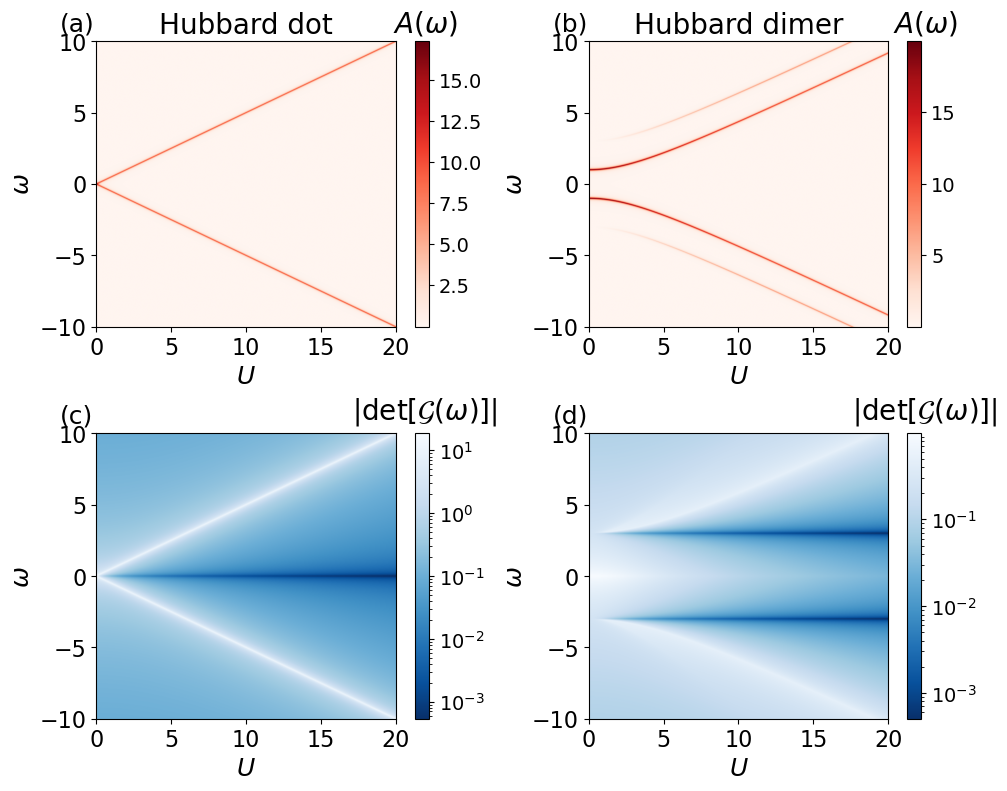}
    \caption{
        \textbf{Poles and zeroes of the Hubbard dot (a,c) and dimer models (b,d).} (a,b): spectral function $A(\omega) = -\frac{1}{\pi}\text{Im} \mathcal{G}(\omega)$, and (c,d): determinant of the single-particle Green's function $|\det [\mathcal{G}(\omega)] |$. Presented as a function of the interaction strength $U$. For the Hubbard dimer, the hopping is set to $t=1$.
    }
    \label{fig:dot_dimer}
\end{figure}

These two operators correspond to distinct charge configurations with different energies. We consider a paramagnetic ground state $|\Psi_0\rangle$ at half-filling, that is $\langle \Psi_0 |{n}_{1\sigma}| \Psi_0 \rangle = \tfrac12$. When $U=0$, the four states in the Hilbert space are degenerate. Computing the GF using Eq.~(\ref{eq:lehmann}), we find that it has a single pole at $\omega=0$. Consequently, and as can be seen directly from the explicit representation of the GF, $\mathcal{G}(\omega)$ is never zero. Moreover, there is no self-energy, so it cannot diverge.
For $U\ne 0$, the eigenstates of the Hubbard dot have different energies: $E^0 = E^{\uparrow\downarrow} = -\tfrac U4$, $E^\uparrow = E^\downarrow = \tfrac U4$. Consequently, the GF at half-filling becomes a sum of two poles separated by $U$:
\begin{align}
    \mathcal{G}(\omega) = \dfrac{1/2}{\omega+\tfrac U2+i0^+}+\dfrac{1/2}{\omega-\tfrac U2+i0^+}.
\end{align}
This GF contains a zero at $\omega=0$, shown in Fig.~\ref{fig:dot_dimer}(c). Writing
\begin{align}
\mathcal{G}(\omega) = \dfrac{1}{\omega-\tfrac{U^2}{4\omega}+i0^+},
\end{align} 
one can see the zero is in correspondence with the divergence of the self-energy $\Sigma(\omega) = \tfrac{U^2}{4\omega}$. In this simple yet important case, the ground-state is degenerate at half-filling as this correspond to a doublet state. This means the spin-excitation are gapless. As a consequence, if the edge state of a topological system in one dimension behave like a Hubbard dot, it is expected to have a correspondence between the GF topology,  with a GF zeroes pinned at $\omega=0$ at the edge and a degenerate ground state under OBC originating from the Hubbard dot degeneracy.

\subsection{Two sites: Hubbard dimer}
A second useful example to consider is the Hubbard Hamiltonian on a dimer, consisting of two Hubbard dots $1$ and $2$ connected by a hopping term (with amplitude $t$):
\begin{align}
    H_{\text{dimer}} = & - t\sum\limits_\sigma\left(c^\dagger_{1\sigma}c_{2\sigma}+c^\dagger_{2\sigma}c_{1\sigma}\right) \\
    \nonumber & + U\sum\limits_{i=1,2}\left(\hat{n}_{i\downarrow}-1/2\right)\left(\hat{n}_{i\uparrow}-1/2\right).
\end{align}
In this case, one can easily compute the GF analytically with Eq.~(\ref{eq:lehmann}). Instead, we present it in Fig.~\ref{fig:dot_dimer}(b,d). One observes that the single-particle GF develops a four-pole structure for $U\ne 0$, reflecting the splitting of the electronic spectral weight induced by interactions. The presence of the hopping term generates off-diagonal (inter-site) components in the GF matrix. Diagonalizing this matrix for $U=0$ separates the excitations into bonding and anti-bonding sectors.

When $U\ne 0$, the anti-bonding states are not eigenstates of the interaction $U$, which leads to some mixing between the two states such that the anti-bonding components acquire finite spectral weight. This gives rise to two GF zeroes located at $\omega = \pm 3t$, which do not disperse with $U$.
In contrast to the single-site case, hopping couples the two local moments and generates an effective antiferromagnetic exchange interaction. As a result, the half-filled ground state is a non-degenerate singlet, separated from the triplet excitations by a finite spin gap
\begin{equation}
    \Delta_s = E_s - E_t
    = \frac{U}{2}- \sqrt{\tfrac{U^2}{4}+4t^2} \xrightarrow[t/U\to 0]{} -\frac{4t^2}{U}.
\end{equation}

\subsection{Additional notes}
The single-site and two-site Hubbard models illustrate a clear correspondence between GF zeroes and local spin physics. When a free spin-\textonehalf{} degree of freedom is present, as in the Hubbard dot, the GF exhibits a zero pinned at zero frequency. When spins are paired into a singlet, as in the Hubbard dimer, the spin gap is finite and the GF zeroes are shifted to finite frequencies.

Breaking particle–hole symmetry, for instance by adding an on-site potential, trivially shifts the position of the zero away from $\omega=0$ without lifting the underlying spin degeneracy. Such shifts do not affect the physical interpretation of the zero as a signature of an emergent local moment and will not be considered further here (see Ref.~\citenum{aligia2025probing}).

These elementary examples provide the conceptual foundation for interpreting boundary Green’s-function zeroes in interacting topological lattice systems.

\section{Effective Hamiltonian for the zeroes}

\subsection{Quasiparticle interpretation\label{Sec:QuasiparticleInterpretation}}

\begin{figure}[h!]
    \centering
    \includegraphics[width=\linewidth]{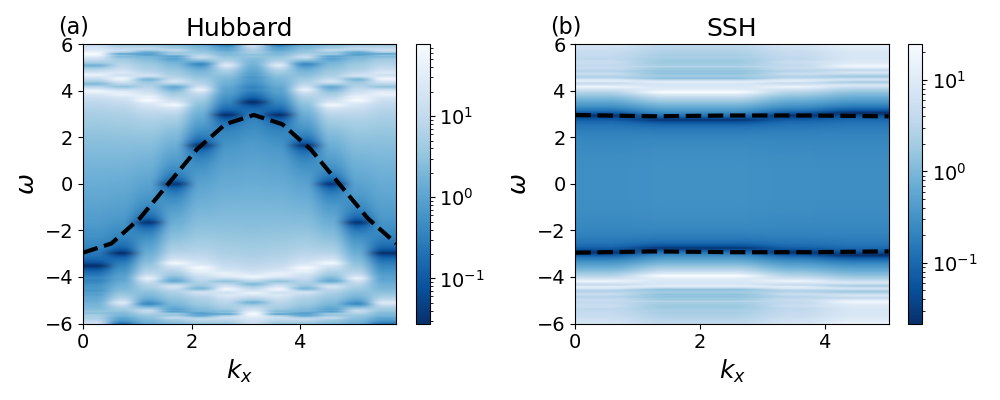}
    \caption{\textbf{Zeroes of the one-dimensional a) Hubbard and b) SSH chains.} We plot the determinant of the electronic GF, $|\det \mathcal{G}(k_x,\omega)|$, computed using exact diagonalization, for one-dimensional (a) Hubbard chain and (b) SSH chains, for systems of size $L=12$. The dispersion of the Green’s function zeroes obtained using the Hubbard-operator method is overlaid as a black dashed line. We use $U=10$ and $t=1$ for the Hubbard model, and $t=0.75$ and $\delta t=0.25$ for the SSH model.}
    \label{fig:1d_zeroes}
\end{figure}

As discussed in the main text, the GF zeroes of a Mott insulator at half-filling can be effectively described by a non-interacting Hamiltonian $\tilde{H}_0$. This object governs the dispersion of the zeroes and plays a role analogous to a Bloch Hamiltonian, but for the absence of electronic spectral weight such that the self-energy takes the form
\begin{align}
\Sigma(\mathbf{k},\omega) \propto \left(\omega - \tilde{H}_0(\mathbf{k})\right)^{-1}.
\end{align}
This representation follows directly from the structure of the Mott insulating state. In this regime, the spectral weight splits into upper and lower Hubbard bands, reflecting an underlying two-pole structure. Consequently, only a single band of zeroes appears within the gap, as shown for the Hubbard model in Fig.~\ref{fig:1d_zeroes}(a). Since these zeroes correspond to poles of the self-energy, they can be captured by an effective non-interacting Hamiltonian describing their dispersion. We emphasize that this construction is specific to half-filling. Away from half-filling, the self-energy acquires a more intricate structure~\cite{sakai2010doped}, and a simple single-band effective Hamiltonian is no longer sufficient to describe its poles. In the doped regime, this formulation must be generalized: the self-energy can no longer be represented as the GF of a non-interacting Hamiltonian, but instead takes the form
\begin{align}
\Sigma(\mathbf{k},\omega) \propto \left(\omega - \tilde{H}_0(\mathbf{k}) - \tilde{\Sigma}(\mathbf{k},\omega)\right)^{-1},
\end{align}
where $\tilde{\Sigma}(\mathbf{k},\omega)$ denotes an additional self-energy associated with the dynamics of the GF zeroes themselves. Since our focus is on the half-filled case and its topological properties, we do not study this case further.

In Refs.~\cite{wagner2023mott,lehmann2025probing,pangburn2025topological}, it was shown that analytical expression for the GF zeroes at half filling can be obtain. The simplest realization of the zeroes Hamiltonian $\tilde{H}_0$ is provided by the two-dimensional Hubbard model treated within a paramagnetic approximation. At half filling, the first-order Hubbard-operator (HO) expansion  (valid in the strong coupling regime) leads to~\cite{pangburn2025topological},
\begin{align}
    & \tilde{H}_0(\mathbf{k})   =2t(1-4p)\epsilon(\mathbf{k}) \\
    & \label{Eq:p_parameter} \quad \text{with} \quad p = \langle n_{i\sigma}n_{j\sigma}\rangle + \langle S^-_iS_j^+\rangle -\langle \Delta_i\Delta_j^\dagger\rangle,
\end{align}
$i$ and $j$ are the two sites connected by the hopping term
where $\Delta_i \equiv c_{i\uparrow}c_{i\downarrow}$, $S^+ \equiv c^\dagger_{i\uparrow} c_{i\downarrow}$, $S^- \equiv c^\dagger_{i\downarrow} c_{i\uparrow}$ and $\epsilon(\mathbf{k})$ the non-interacting dispersion of the original system. This expressions assume translation invariance and agrees with other perturbative expansions~\cite{lehmann2025probing}. The coefficient $p$ encodes the strength of charge, spin, and pairing correlations on the bond $\langle ij \rangle$. 

We observe that the GF zeroes correspond to a renormalized dispersion of the non-interacting band structure. This already suggests that GF zeroes are unable to capture beyond–single-particle physics. For multi-orbital models (or two sites in the unit cell such as the SSH chain), the structure of the zeroes Hamiltonian generalizes to
\begin{align}
    \tilde{H}_{0,\alpha\beta}(\mathbf{k}) & =2t_{\alpha\beta}\left(1-4p_{\alpha\beta}\right)\gamma_{\alpha\beta}(\mathbf{k}), \ \text{with} \\
    \nonumber p_{\alpha\beta} & = \langle n_{\alpha\sigma} n_{\beta\sigma}\rangle + \langle S_{\alpha}^-S_{\beta}^+\rangle -\langle \Delta_{\alpha} \Delta_{\beta}^\dagger \rangle.
\end{align}
$\alpha$ and $\beta$ are orbital/layer indices, while $\gamma_{\alpha\beta}(\mathbf{k})$ is the form factor that couples orbital $\alpha$ to the orbital $\beta$.

In Fig.~\ref{fig:1d_zeroes}, we compare the GF zeroes of the one-dimensional Hubbard and SSH chains obtained from exact diagonalization (ED), as the color plot, and from the Hubbard-operator (HO) approximation, as the dashed line. In both models, we find a good agreement between the analytical results and the exact numerical data. This agreement validates the use of the zeroes Hamiltonian $\tilde{H}_0$ as a reliable tool for analyzing GF zeroes in Mott insulators. For a greater accuracy, one can systematically improve $\tilde{H}_0$ by going to the next order of the HO approximation.

\subsection{Winding number of $\tilde{H}_0(\mathbf{k})$\label{App:Winding_HO}}
The zeroes Hamiltonian respects the same symmetries as the non-interacting Hamiltonian. For the SSH chain, in the sublattice basis, $\tilde{H}_0(\mathbf{k})$ read 
\begin{align}
& \tilde{H}_0(\mathbf{k}) = 
    \begin{pmatrix}
        0 & q_0(\mathbf{k}) \\
        q^\dagger_0(\mathbf{k}) & 0 \\
    \end{pmatrix}&\\[5pt]
&q_0(\mathbf{k}) =t_+(1-4p_+) e^{\tfrac{i\mathbf{k}}2} + t_-(1-4p_-)e^{-\tfrac{i\mathbf{k}}2}&
\end{align}
with $t_+ = t+\delta t$ and $t_-=t-\delta t$. The expression of $p_\pm$ is given in Eq.~(\ref{Eq:p_parameter}), where $+$ ($-$) is taken along the stronger (weaker) bond. The winding of the GF zeros is given by Eq.~(\ref{eq:NI_nu}) for $q_0(\mathbf{k})$.
For the parameters $t$ and $\delta t$ considered in this work, $t_+(1-4p_+) \neq t_-(1-4p_-)$ for all $U$ which implies the gap of $\tilde{H}_0(\textbf{k})$ is not closing as a function of $U$. Consequently, the topological invariant associated with the zeroes, $\nu$, remains constant throughout the interaction range.

\subsection{Weakly interacting regime}
GF invariants were originally proposed for weakly interacting regimes. However, even in this limit, they fail to probe topology correctly. Indeed, for sufficiently weak interactions, the many-body gap does not close upon turning on interactions for PBC. Similarly, although zeroes of the GF can appear for infinitesimal interactions as illustrated for the Hubbard dot and dimer, the GF invariants remain unchanged in the absence of a poles or zeroes gap closing when the interaction strength is small enough. In contrast, the nature of the edge states can change drastically, even for arbitrarily weak interactions, by acquiring a finite gap. Consequently, in the weakly interacting regime, a discrepancy arises between GF bulk invariants, which remain unaffected, and edge properties, which can undergo qualitative changes. For instance, two coupled SSH chains are fully trivial in the limit $U\rightarrow 0$, $J_{12}\rightarrow 0$ since any infinitesimal $U$ lifts the degenerate charge mode at the boundary, while any infinitesimal $J_{12}$ removes the remaining spin degeneracy.

\section{Numerical methods}

Throughout this work, we employ various numerical methods detailed in this section. 

\paragraph{Exact diagonalization.---}
In order to compute the real-frequency GF accurately, we need to compute not only the ground state, but also the full spectrum of excitations. For that, we are using Lanczos method~\cite{gagliano1988dynamic, dagotto1994correlated}. This allows us to access system size, up to $L=12$ with spinful electrons in this work. It was benchmark with the Lehmann representation result for $L=8$ taking into account all states.

For larger systems which are inaccessible with this method, we employ the Hubbard operators (HO) method, detailed in Sec.~\ref{App:HubbardOperatorFormalism}. GF zeros within the HO method    are presented in Sec.~\ref{App:HO_GF_zeroes}, and benchmarked are shown in Sec.~\ref{App:benchmark_HO}.

\paragraph{Density-matrix renormalization group.---}
For the one-dimensional gapped systems studied in the work, the density-matrix renormalization group (DMRG) is optimal to compute the ground state of large systems and fully capture the topological properties~\cite{white1992density,schollwock2005density, SCHOLLWOCK201196}.
We used the ITensor library~\cite{Itensor}, and calculations were performed employing a cutoff of $\epsilon = 10^{-10}$, a maximum bond dimensions of $\chi  = 1024$, and sweeps were performed until convergence of the energy down to change of $10^{-6}$. For the two-dimensional system, we used a bond dimension of $\chi = 6000$.

\paragraph{Auxiliary-field quantum Monte Carlo.---}
For the two-dimensional system studied in this work (the interacting BBH layers), we 
employed auxiliary-field quantum Monte Carlo (AFQMC)~\cite{zhang1997constrained,zhang2003quantum}, using the SAFIRE software~\cite{safire26}. We used
a Trotter time step of $\Delta\tau=0.005$ and we performed $24 000$ time steps to propagate the walkers, with a total walker population of $N_w = 3200$. All simulations employ a real auxiliary-field bias and the constrained-path propagator. We compared the ground-state energy between AFQMC with those parameters and DMRG on cylindrical geometries, presented in Fig.~\ref{fig:afqmc_vs_dmrg}. We find that the agreement is sufficient and within statistical error bars, in particular the low-$U$ regime.

\paragraph{Spin and charge gaps.---}

The spin and charge gaps $\Delta_s$ and $\Delta_c$ are defined from differences between many-body ground-state energies in different quantum-number sectors. Because we consider models with $U_\uparrow(1)\times U_\downarrow(1)$ symmetry, one can compute ground-state energies for fixed number of particles, $N_\uparrow$ spins up and $N_\downarrow$ spins down, noted $E_0(N_\uparrow,N_\downarrow)$. The gaps are then defined as
\begin{align}
\Delta_s & = E_0(N_\uparrow+1,N_\downarrow-1)-E_0(N_\uparrow, N_\downarrow)\\
\Delta_c & = E_0(N_\uparrow+1,N_\downarrow)+E_0(N_\uparrow-1,N_\downarrow)-2E_0(N_\uparrow,N_\downarrow). \nonumber
\end{align}

\section{Hubbard operator method\label{App:HubbardOperatorFormalism}}
The Hubbard operator (HO) method employed in this manuscript is based on solving exactly some finite clusters, then coupling these clusters together. Conceptually, the method extends standard mean-field theory by incorporating correlation functions beyond those retained in a conventional mean-field decoupling. It is exact in two limits: the non-interacting limit ($U=0$) and the fully atomic limit ($t=0$), and it is expected to be accurate in the vicinity of these two regimes. However, for $t \sim U$, there are no general arguments to guarantee its validity.

In this section, we first thoroughly present the  method in App.~\ref{app:HO_formalism}. We then apply the method to solve for the GF zeroes of various systems in App.~\ref{App:HO_GF_zeroes}. In App.~\ref{App:benchmark_HO}, we present several benchmarks to assess the performance of the HO method in the intermediate regime. These benchmarks support its reliability in the models and across their whole range of parameters employed in this work.

\subsection{Formalism}
\label{app:HO_formalism}
The HO method~\cite{hubbard1963electron,hubbard1964electron,roth1969electron,mancini2004hubbard} starts by decomposing the Hamiltonian into local and non-local parts,
\begin{align}
H = \sum_i H_i^{\text{loc}} + \sum_{ij} H^t_{ij},
\end{align}
where $H_i^{\text{loc}}$ acts on a finite-dimensional local Hilbert space $\mathcal{H}_i$, and $H^t_{ij}$ couples different sites $i$ and $j$.

Because $\mathcal{H}_i$ is finite-dimensional, one can construct a finite operator basis $\mathbf{\Psi}_i^{\text{loc}}$ that closes under the equation of motion with the local Hamiltonian:
\begin{align}
\mathbf{J}_{i,n}^{\text{loc}} \equiv \left[\boldsymbol{\Psi}_{i,n}^{\text{loc}}, H_i^{\text{loc}}\right]
= \sum\limits_{m}E_{i,nm}^{\text{loc}}\boldsymbol{\Psi}_{i,m}^{\text{loc}},
\end{align}
where $E_i^{\text{loc}}$ is a matrix whose eigenvalues describe the local excitation energies and $[\cdot,\cdot]$ is the commutator. In the following, the indices $i$, $j$ and further $l$ denote site indices, while $n$, $m$ and further $p$ label the operators in the chosen basis.
Although we focus on charge excitations, the same construction applies to operators with different quantum numbers, such as spin excitations~\cite{mancini2004hubbard}.

The hopping term generates additional contributions to the equations of motion,
\begin{align}
\Big[\boldsymbol{\Psi}_{i,n}^{\text{loc}}\ ,\sum\limits_j H^t_{ij}\Big] = \delta \mathbf{J}_{i,n} \ ,
\end{align}
where $\delta\mathbf{J}_i$ contains operators that cannot be expressed as linear combination of operators present in $\bigoplus_l \bm \Psi^{\text{loc}}_l$. 
The key approximation of the HO method is to assume that the chosen basis already captures the relevant quasiparticles of the problem. One therefore projects $\delta\mathbf{J}_i$ back onto $\bigoplus_l \bm \Psi^{\text{loc}}_l$ :
\begin{align}
   \delta \mathbf{J}_{i,n} \approx \sum\limits_m\sum\limits_{j}E^t_{ij,nm} \boldsymbol{\Psi}_{j,m}^{\text{loc}}.
\label{eq:OperatorBasis}
\end{align}
$E^t$ is the linear approximation corresponding to the projection of $\delta\mathbf{J}_i$ to $\bigoplus_l \bm \Psi^{\text{loc}}_l$. The nature of the quasiparticles is thus fixed by the local Hamiltonian, while $H^t$ couples them perturbatively.

Combining the local and non-local contributions yields
\begin{align}
\mathbf{J}_{i,n} \approx \mathbf{J}_{i,n}^{\text{loc}} + \delta\mathbf{J}_{i,n}
= \sum\limits_m\sum\limits_{j} \underbrace{(\delta_{ij}E^{\text{loc}}_i + E^t_{ij})_{nm}}_{E_{ij,nm}}\boldsymbol{\Psi}_{j,m}^{\text{loc}}.
\label{eq:E_matrix_def}
\end{align}
Evaluating $E$ requires two matrices defined by expectation values:
\begin{align}
    \label{eq:HO_M_and_I_matrices}
    M_{ij,nm} & = \langle\left\{ {\mathbf{J}_{i,n}, \boldsymbol{\Psi}_{j,m}^{\text{loc},\dagger}} \right\}\rangle, \\
    I_{ij,nm} & = \langle \left\{ \boldsymbol{\Psi}_{i,n}^{\text{loc}}, \boldsymbol{\Psi}^{\text{loc},\dagger}_{j,m} \right\}\rangle,
\end{align}
where all anti-commutators are evaluated at equal times. The expectation value $\langle \ldots\rangle$ denotes either a ground-state average or a thermal average, depending on the context. 
The elements of the quasiparticle energy matrix are given by~\cite{haurie2024bands,mancini2004hubbard}
\begin{align}
    \label{eq:HO_E_matrix}
    E_{ij,nm} =\sum\limits_p\sum\limits_l M_{il,np} (I^{-1})_{lj,pm}.
\end{align} 
Both $M$ and $I$ depend on expectation values that must be determined self-consistently, so the method naturally becomes a self-consistent scheme, more involved than in standard mean-field theory.
In this work, we employ the Roth decoupling scheme~\cite{roth1969electron}, recently reviewed in Refs.~\cite{haurie2024bands, banerjee2025charge}. The decoupling allows to compute higher-order correlations function like $\langle S_i^+S_j^- \rangle$ that cannot be computed directly from the single-particle GF.

The retarded GF for the operator basis is
\begin{align}
    \mathds{G}_{ij,nm}(t)
    = -i\theta(t)\langle \left\{ \boldsymbol{\Psi}^\text{loc}_{i,n}(t),
    \boldsymbol{\Psi}^{\text{loc},\dagger}_{j,m}(0) \right\}\rangle .
\end{align}
Using Eq.~\eqref{eq:E_matrix_def}, one obtains in frequency space:
\begin{align}
    \mathds{G}_{ij,nm}(\omega)
    = \left((\omega+i\eta - E)^{-1}I\right)_{ij,nm},
    \label{eq:GF_def}
\end{align}
where $\eta$ is a infinitesimal quasiparticle broadening.
The presence of the non-canonical matrix $I$ (defined in Eq.~(\ref{eq:HO_M_and_I_matrices}) is what distinguishes this interacting theory from a free-fermion model, as it encodes the non-canonical algebra obeyed by the operators.

When including in the operator basis holons $\xi_{i\sigma}$ and doublons $\eta_{i\sigma}$, defined by
\begin{align}
    \xi_{i\sigma} = c_{i\sigma}(1-n_{i\bar\sigma}),\qquad
    \eta_{i\sigma} = c_{i\sigma}n_{i\bar\sigma}
\end{align}
with $\bar{\sigma} = - \sigma$,
the electron operator decomposes as
\begin{align}
    c_{i\sigma} = \xi_{i\sigma} + \eta_{i\sigma}.
\end{align}
The electronic retarded GF $\mathcal{G}$ can then be expressed as:
\begin{align}
    \mathcal{G}_{ij}
    = \mathds{G}_{ij, \xi\xi}+\mathds{G}_{ij, \eta\xi}+\mathds{G}_{ij, \xi\eta}+\mathds{G}_{ij,\eta\eta}.
\end{align}

\subsection{Green's function zeroes from Hubbard operators\label{App:HO_GF_zeroes}}

Because of the analytical structure of the HO formalism, explicit expressions for the GF zeroes can be obtained directly by solving the equation 
\begin{align}
\det \mathcal{G}(\omega)=0.
\end{align}
In the following, we present the solution of this equation in some simple models, which lead to the definition of the zeroes Hamiltonian presented in Sec.~\ref{Sec:QuasiparticleInterpretation}.

\subsubsection{Single-orbital Hubbard model}

For the single band Hubbard model, the dispersion of GF zeroes at a filling $n=1$ can be readily obtained from Ref.~\citenum{pangburn2025topological}. At half-filling, the $M$-matrix defined in Eq.~(\ref{eq:HO_M_and_I_matrices}) is block diagonal with translation symmetry. It is given by
\begin{align}
&M(\mathbf{k})  =\begin{pmatrix}
-\frac{U}{2}-\gamma(\mathbf{k}) p & -\gamma(\mathbf{k})(\frac{1}{2}-p) \\
-\gamma(\mathbf{k})(\frac{1}{2}-p) & \frac{U}{2}-\gamma(\mathbf{k}) p \\
\end{pmatrix} \\
  &  \quad \text{where} \quad \gamma(\mathbf{k}) = 2t\left(\cos(k_x)+\cos(k_y)\right),
\end{align}

with $p$ already defined in Eq.~\ref{Eq:p_parameter}. The $I$-matrix also defined in Eq.~(\ref{eq:HO_M_and_I_matrices}) is very simple, given by $I = \frac{1}{2}\mathds{1}_2$. 

Using that $\mathcal{G}(\mathbf{k},\omega)=\sum_{\alpha,\beta=1}^2\mathds{G}_{\alpha\beta}(\mathbf{k},\omega)$, we can find each frequencies $\omega_0(\textbf{k})$ such that $\det \mathcal{G}\left(\mathbf{k},\omega_0(\textbf{k})\right)=0$. For each $\mathbf{k}$, there is a single solution $\omega_0(\mathbf{k})$ given by the following expression:
\begin{align}
 \omega_0(\mathbf{k})&=\left(M_{11}(\mathbf{k})+M_{22}(\mathbf{k})-2 M_{12} \nonumber (\mathbf{k})\right) \label{Eq:zeroesSingle} \\
 &=-2t(1-4p)(\cos(k_x)+\cos(k_y)).
\end{align}
The zeroes are simply proportional to the non-interacting dispersion and become completely dispersionless at $p=\frac{1}{4}$, which corresponds to the Hubbard-I approximation~\cite{hubbard1963electron}.

\subsubsection{General case}
For an arbitrary complicated model, a formula can be found for $\tilde{H}_0$ in the first order HO approximation, as was shown in Ref.~\cite{pangburn2025topological}.
In particular, for a half-filled Mott insulator, the electronic GF zeroes are given by the following Hamiltonian:
\begin{align}
\label{Eq:zeroesPrediction}
    \tilde{H}_{0, \alpha\beta}(\mathbf{k}) &= \left[M_{\alpha\beta}(\mathbf{k})\right]_{\xi\xi}+\left[M_{\alpha\beta}(\mathbf{k})\right]_{\eta\eta}&\\
    &-\left[M_{\alpha\beta}(\mathbf{k})\right]_{\xi\eta}-\left[M_{\alpha\beta}(\mathbf{k})\right]_{\eta\xi}. \nonumber
\end{align}
where $\alpha$ and $\beta$ are orbital indices of the $M$ matrix and translation invariance is assumed.

\subsubsection{Spin-spin coupling\label{App:SpinSpinCoupling}}

We now consider a Hubbard dimer including both hopping and spin–spin interactions. The dimer Hamiltonian reads
\begin{align}
\nonumber
H & =U\sum_{i\in\{a,b\}}\left(\hat{n}_{i\uparrow}-\tfrac{1}{2}\right)\left(\hat{n}_{i\downarrow}-\tfrac{1}{2}\right)-t\sum\limits_{\sigma}\left(c^\dagger_{a\sigma}c_{b\sigma}+h.c.\right) \\
 & +J\left(S_{x,a}S_{x,b}+S_{y,a}S_{y,b}+S_{z,a}S_{z,b}\right)
\end{align}
where $a$ and $b$ are the two dimer sites.

Introducing the composite operator basis
\begin{align}
    \label{eq:basis_general_case}
    \psi_\uparrow = \left(\xi_{a\uparrow},\eta_{a\uparrow},\xi_{b\uparrow},\eta_{b\uparrow}\right)^T,
\end{align}
the $M$-matrix in this basis reads
\begin{align}
M= \begin{pmatrix}
M_{aa} & M_{ab} \\
M_{ba} & M_{bb} \\
\end{pmatrix}.
\end{align}
$M_{aa}$ and $M_{bb}$ are intra-site components, while $M_{ab}$ and $M_{ba}$ are inter-site components. By symmetry, $M_{ij} = M_{ji}$.

The intra-site elements can be shown to be
\begin{align}
    M_{aa,\xi\xi}  = & \ \tfrac J4\big(\langle n_{a\downarrow}n_{b\downarrow}\rangle -\langle n_{a\downarrow}n_{b\uparrow}\rangle + 2\langle S_a^+S_b^-\rangle\big)  \\
    \nonumber & \  - \tfrac U2\big(1-\langle n_{a\downarrow} \rangle\big), \\
    M_{aa, \eta\eta}  = & \ - \tfrac J4\big(\langle n_{a\downarrow}n_{b\downarrow}\rangle -\langle n_{a\downarrow}n_{b\uparrow}\rangle + 2\langle S_a^+S_b^-\rangle\big)  \\
    \nonumber & \ + \tfrac U2\langle n_{a\downarrow}\rangle, \quad \quad \text{and} \\
M_{aa, \xi\eta} = & \ M_{aa, \eta\xi} = 0,
\end{align}
while the inter-site elements are found as
\begin{align}
    \nonumber M_{ab, \xi\xi} = &  -t \big(1-\langle n_{a\downarrow}\rangle-\langle n_{b\downarrow}\rangle+p_{ab}\big) \\
        &+ \tfrac {3J}4 \langle \xi_{a\sigma}\xi^\dagger_{b\sigma}\rangle,  \\
    M_{ab,\xi\eta}= & -t\left(\langle n_{a\downarrow}\rangle-p_{ab}\right) + \tfrac {3J}4\langle \xi_{a\sigma}\eta^\dagger_{b\sigma}\rangle,   \\
    M_{ab,\eta\xi} = & -t\left(\langle n_{b\downarrow}\rangle-p_{ab}\right) + \tfrac{3J}4\langle \eta_{a\sigma}\xi^\dagger_{b\sigma}\rangle, \\
    \text{and} & \quad M_{ab,\eta\eta} = -tp_{ab} + \tfrac{3J}4\langle \eta_{a\sigma}\eta^\dagger_{b\sigma}\rangle,  \\
    \text{where} & \quad p_{ab} = \langle n_{a\uparrow}n_{b\uparrow}\rangle + \langle S_a^-S_b^+\rangle -\langle \Delta_a\Delta_b^\dagger\rangle.
\end{align}

Eq.~(\ref{Eq:zeroesPrediction}) can then be used to derive an analytical expression for the GF zeroes in the presence of spin–spin interactions. Following Sec.~\ref{App:HO_GF_zeroes}, the zeroes Hamiltonian of the dimer reads
\begin{align}
&\tilde{H}_0 = \begin{pmatrix}
M^{aa}_{\xi\xi}+M^{aa}_{\eta\eta}-2M_{\xi\eta}^{aa} & M^{ab}_{\xi\xi}+M^{ab}_{\eta\eta}-2M_{\xi\eta}^{ab}\\
M^{ba}_{\xi\xi}+M^{ba}_{\eta\eta}-2M_{\xi\eta}^{ba} & M^{bb}_{\xi\xi}+M^{bb}_{\eta\eta}-2M_{\xi\eta}^{bb} \\
\end{pmatrix},
\end{align}
Equation~(\ref{Eq:zeroesPrediction}) can then be simplified to
\begin{align}
    \tilde{H}_{0,11} = \tilde{H}_{0,22} & = 0 \\
    \tilde{H}_{0,12} = \tilde{H}_{0,21} & = -t \big(1-4p_{ab}\big) \\
    & + \tfrac{3J}4 \left(\langle \xi_{a\sigma}\xi_{b\sigma}^\dagger\rangle + \langle\eta_{a\sigma}\eta_{b\sigma}^\dagger\rangle-2\langle \xi_{a\sigma}\eta^\dagger_{b\sigma}\rangle\right). \nonumber
\end{align}
Inserting $J=0$, one can also reproduce the Hubbard operator results presented in Fig.~(\ref{fig:dot_dimer}).

\subsection{Benchmarking}
\label{App:benchmark_HO}

\subsubsection{Results}

In this section, we apply the Roth-decoupling scheme to the models studied in this work to check the validity of the HO method.

\begin{figure}[h!]
    \centering
    \includegraphics[width=1.0\linewidth]{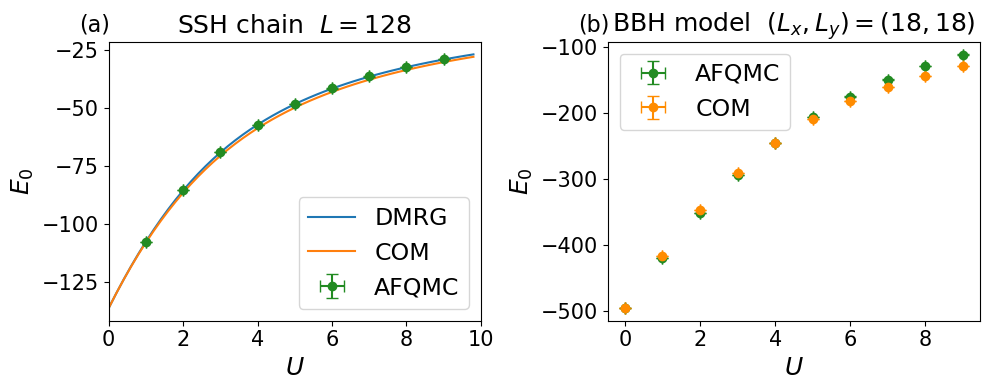}
    \caption{\textbf{Ground-state energy benchmark at half-filling} for (a) the SSH and (b) the BBH models, obtained with HO, DMRG and AFQMC. (a) The ground-state energy is shown as a function of $U$ with parameters $t=0.75$, $\delta t = 0.25$ with open boundary condition with trivial boundary mode for a size $L=128$. (b) The same quantity for the BBH model in a topologically trivial state with open boundary condition for a lattice of size $(L_x,L_y) = (18,18)$ with parameters $\lambda_x = \lambda_y = 0.5$ and $\gamma_x = \gamma_y = 1.0$. }
    \label{fig:benchmark_COM_GS_energy}
\end{figure}

\begin{figure}
    \centering
    \includegraphics[width=1.0\linewidth]{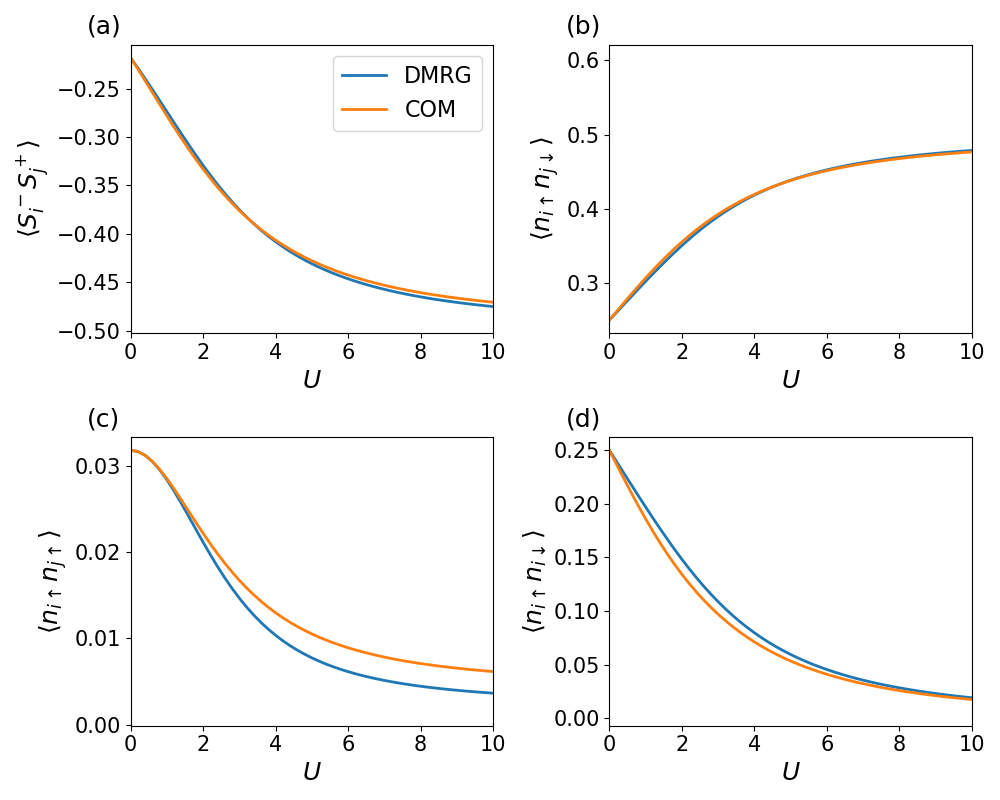}
    \caption{\textbf{Correlation-function benchmark at half filling} for the SSH chain with parameters $t=0.75$ and $\delta t=0.25$, computed for a system size $L=128$. Results obtained using the HO approximation (orange lines) are compared to DMRG calculations (blue lines). Shown are bulk correlation values evaluated on the strong bonds of the chain, including the transverse spin correlation $\langle S_i^- S_j^+ \rangle$ (a), the spin-resolved density correlation $\langle n_{i\uparrow} n_{j\downarrow} \rangle$ (b), the same-spin density correlation $\langle n_{i\uparrow} n_{j\uparrow} \rangle$ (c), as well as the local double occupancy $\langle n_{i\uparrow} n_{i\downarrow} \rangle$ (d).}
    \label{fig:benchmark_COM_corr_function}
\end{figure}

In Fig.~\ref{fig:benchmark_COM_GS_energy} we provide ground-state energy benchmark for the SSH and BBH model. For the former, a one-dimensional model, the HO method yields a result that is already very close to the exact solution provided by DMRG, for a large range of values for the Hubbard repulsion $U$ parameter. Note that because the method is not variational, the ground-state energy can be lower than the exact result, which is the case here. For the two-dimensional BBH model, the HO method gives less accurate results in particular at large value of $U$, in comparison with AFQMC. This is expected for two reasons. First, the hopping perturbation is larger as each site has more neighbor than the 1D case. Second, at large $U$, some antiferromagnetic order sets in the BBH model, which the HO method is not capturing in the paramagnetic approximation.

In Fig.~\ref{fig:benchmark_COM_corr_function}, we present a benchmark of correlation functions computed using the Hubbard operator method with the Roth decoupling, compared to DMRG results. The method is exact at $U=0$ and remains accurate over the entire interaction range at half filling. Unlike mean-field approaches, which are limited to single-particle expectation values, the Roth decoupling scheme enables the calculation of selected two-particle correlation functions. For the operator basis considered here, it accurately captures the following computed correlators: $\langle n_{i\uparrow}n_{j\uparrow}\rangle$, $\langle n_{i\uparrow}n_{j\downarrow}\rangle$, $\langle S_i^-S_j^+\rangle$ and $\langle \Delta_i\Delta^\dagger_j\rangle$ for nearest-neighbor sites $i$ and $j$, as these explicitly enter the decoupling scheme. Correlation functions beyond this set cannot be reliably computed within the holon–doublon operator basis and require an enlarged basis.

\section{SSH chain}

In this section, we present additional results for the SSH chain. In App.~\ref{app:ssh_isolated}, we show
\begin{enumerate}
    \item The GF topology as a function of $U$. Starting form a topological states, the GF poles are trivialized and the GF zeros keep the topolog of the non-interacting state.
    \item The evolution of the ground-state topology as a function of $U$ which is directly connected to the non-interacting topology.
\end{enumerate}
which support some of the main text claims.
In App.~\ref{app:ssh_two_chains}, we provide additional results for the coupled chains. In particular we compare the results obtained by the HO method with ED, and that the zeroes are gapped under PBC.

\subsection{Isolated chain}
\label{app:ssh_isolated}
In the SSH chain, the distinction between the topological and trivial phases is geometric and is controlled by the pattern of alternating hopping amplitudes. In the topological phase, the chain terminates on weak bonds, leaving an unpaired site at each boundary equivalent to a Hubbard dot, whereas in the trivial phase the chain ends on strong bonds, forming a boundary dimer.

We compute the topological invariant of the SSH chain using the zeroes Hamiltonian as defined in Sec.~\ref{App:Winding_HO}. Within this framework, we find that the interacting invariant $\nu[\tilde{H}_0]$ in Eq.~(\ref{eq:NI_nu}) remains identical to its non-interacting value ($U=0$) across the full range of interaction strengths considered ($\nu[\tilde H_0] = \pm 1$). The mechanism underlying the robustness of the invariant with respect to $U$ can be understood as follows. When interactions are introduced, additional poles associated with the upper and lower Hubbard bands acquire finite spectral weight, whereas in the non-interacting limit these poles carry vanishing weight. The only way for such poles to disappear continuously as $U \to 0$ is through cancellation with zeroes of the Green’s function, which indeed occurs. Since, for $U \neq 0$, these poles and zeroes of the single-particle GF carry the same topological character but contribute with opposite signs to GF based topological invariants, their mutual cancellation leaves the value of the invariant unchanged for any $U$. This result is consistent with a direct computation of the Berry phase via spin-polarized flux insertion~\cite{berry1984quantal,xiao2010berry}.

\begin{figure}[h!]
    \centering
    \includegraphics[width=\linewidth]{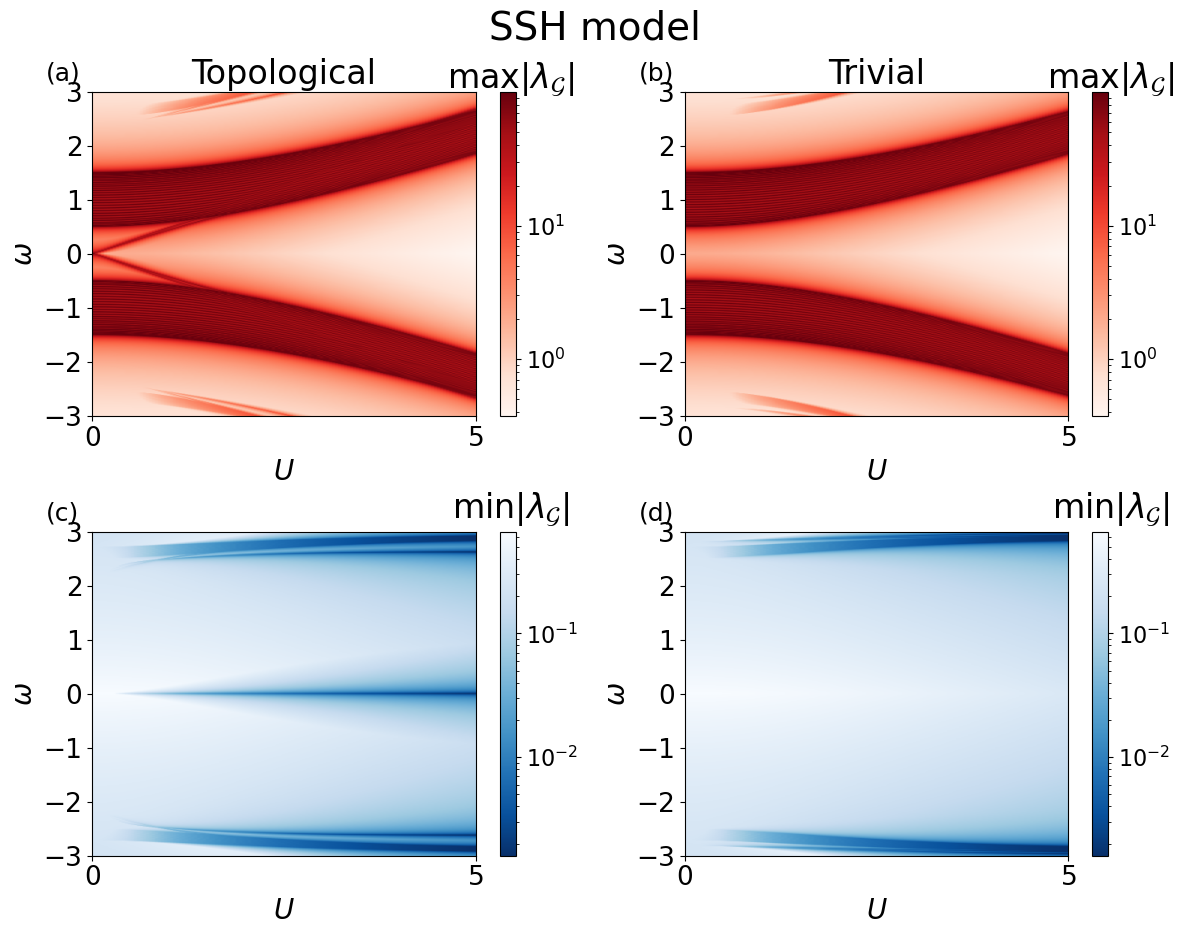}
    \caption{
        \textbf{Poles and zeroes of the topological (a,c) and trivial (b,d) SSH model.}
        Maximum (a,b) and minimum (c,d) eigenvalues of the interacting GF, corresponding respectively to poles and zeroes, for an SSH chain with open boundary conditions and length $L=64$ as a function of the Coulomb repulsion $U$. They are obtained with the Hubbard operator method (HO). (a,c): topological SSH chain with $t_1=1$ and $t_2=$~\textonehalf. (b,d): trivial SSH chain with $t_1=$~\textonehalf \ and $t_2=1$.
    }
    \label{fig:poles_zeroes_SSH}
\end{figure}

\begin{figure}[ht!]
    \centering
    \includegraphics[width=\linewidth]{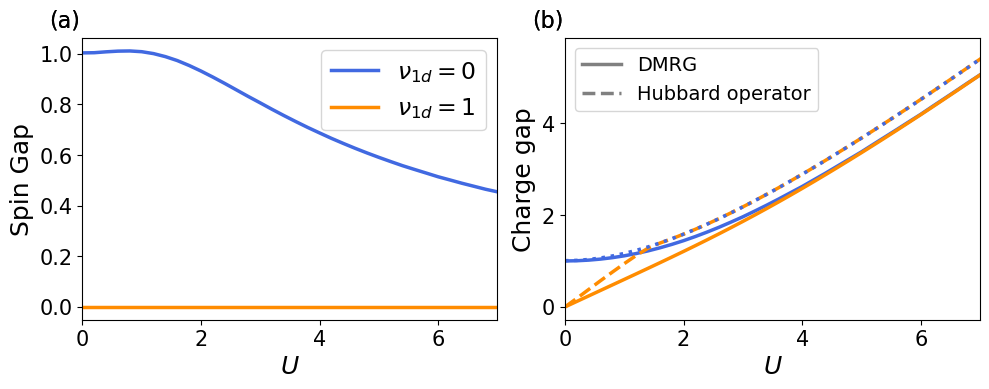}
    \caption{\textbf{Spin (a) and charge (b) gaps for a SSH chain of size $L=128$ with open boundary condition as a function of Coulomb repulsion $U$.}
    Solid lines represent DMRG results, whereas dotted lines indicate calculations performed using the Hubbard operator method (HO). The trivial SSH chain ($\nu_{1d}=0$) is shown in blue, and the topological SSH chain ($\nu_{1d}=1$) in orange. The hopping parameters are fixed to $t=0.75$ and $\delta t = 0.25$.
    }
    \label{spin_charge_gaps_SSH}
\end{figure}

\paragraph{GF topology as a function of $U$ : poles trivialization and topological zeros}

In Fig.~\ref{fig:poles_zeroes_SSH}, we present the boundary GF spectrum of the SSH chain with open boundary conditions in both the topological and trivial phases. We study the eigenvalue spectrum $\lambda_{\mathcal{G}}(\omega,U)$ of the single-particle GF. This formulation allows poles and zeroes to be tracked on equal footing: poles correspond to divergences in the largest eigenvalue, while zeroes are identified by the vanishing of the smallest eigenvalue.

For $U \neq 0$, the largest eigenvalue $\max[\lambda_{\mathcal{G}}]$ reveals that the zero-energy boundary mode characteristic of the topological phase splits into two peaks at $\omega=\pm U/2$. This behavior is in direct agreement with the spectrum of an isolated Hubbard dot and reflects the effective decoupling of the boundary state from the bulk. The smallest eigenvalue $\min[\lambda_{\mathcal{G}}]$, which probes GF zeroes, simultaneously develops a zero pinned at the boundary as soon as interactions are introduced. This observation is consistent with the nontrivial topological index computed from the zeroes Hamiltonian $\tilde{H}_0$.

In contrast, in the trivial phase both poles and zeroes remain gapped for all values of $U$. This behavior is characteristic of a Hubbard-dimer–like boundary spectrum and reflects the absence of an emergent boundary degree of freedom in this case.

\paragraph{Ground-state topology}

In Fig.~\ref{spin_charge_gaps_SSH}, we show the spin and charge gaps of the SSH chain as a function of $U$ in both the topological and trivial phases, computed using DMRG and compared with HO for the charge gap. In the topological phase, the charge sector becomes gapped immediately for $U>0$, while the spin sector remains gapless, in correspondence with the presence of a boundary GF zero. By contrast, in the trivial phase both charge and spin excitations are gapped for any $U$, which is likewise reflected in the GF spectrum by the absence of zero-energy poles or zeroes. This shows that, in the case of an isolated SSH chain, the topology of the interacting state is directly connected to the non-interacting limit.

\subsection{Two coupled chains}
\label{app:ssh_two_chains}

\begin{figure}[ht!]
    \centering
    \includegraphics[width=\linewidth]{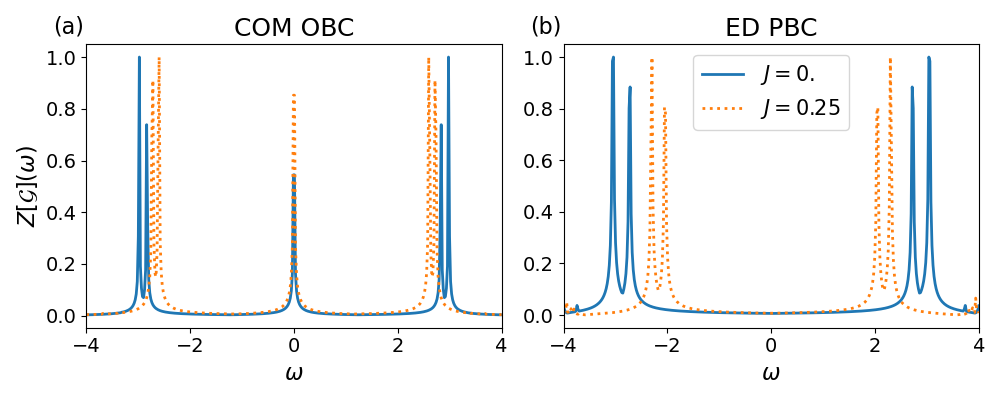}
    \caption{\textbf{Topology inferred from Green's function zeroes in two SSH chains coupled by many-body interactions.} (a)~GF zeroes probe $Z[\mathcal{G}](\omega)$ computed with HO for two SSH chains of length $L=6$ with $U=8$, $t=0.75$ and $\delta t=0.25$ coupled with spin-spin coupling $J_{12}=0$ and $J_{12}=0.25$, with open boundary conditions. (b)~GF zeroes probe $Z[\mathcal{G}](\omega)$ computed with ED for two SSH chains of length $L=6$ with $U=8$, $t=0.75$ and $\delta t=0.25$ coupled with spin-spin coupling $J_{12}=0$ and $J_{12}=0.25$, with periodic boundary conditions.
    }
    \label{fig:two_SSH_chains_J12}
\end{figure}

In this section, we present additional results supporting the coupled-chains topology under coupling. Figure~\ref{fig:two_SSH_chains_J12}(a) shows the GF zeroes invariant computed with HO for the coupled SSH chain with $J_{12}=0$ and $J_{12} = 0.25$ under OBC. These results agree closely with those obtained with ED, presented in the main text, showing the analytics are enough to explain the numerical results. In Fig.~\ref{fig:two_SSH_chains_J12}(b), we show the ED GF zeroes indicator under PBC. As stated in the main text, the GF zeroes gap is not closing for $J_{12}\ne 0$, such that the GF topological invariant is still well-defined for small enough $J_{12}$.

\begin{figure}[ht!]
    \centering
    \includegraphics[width=\linewidth]{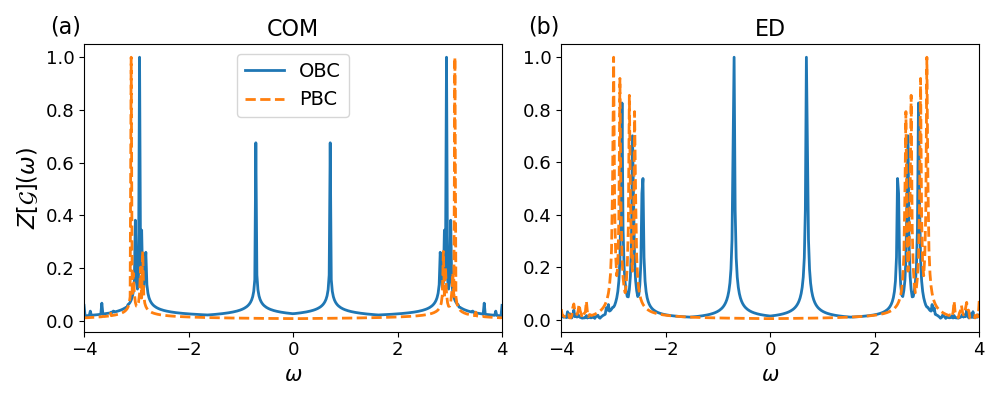}
    \caption{\textbf{Topology inferred from Green's function zeroes in two SSH chains coupled by single-body interactions.} GF zeroes probe $Z[\mathcal{G}](\omega)$ computed with HO (a) and  ED (b) for two SSH chains of length $L=6$ with $U=8$, $t=0.75$ and $\delta t=0.25$ coupled with hybridization interactions $t_{12}=0$ and $t_{12}=0.25$, with open and periodic boundary conditions.
}
\label{fig:two_SSH_chains_t12}
\end{figure}

In Fig.~\ref{fig:two_SSH_chains_t12}, we present the GF zeroes spectrum for two topological SSH chains, described by Hamiltonians $H_1$ and $H_2$, coupled via a single-particle interchain hopping
\begin{align}
H_{12}' = t_{12} \sum_{i,\sigma} \left( c^{\dagger}_{i1\sigma} c_{i2\sigma} + \mathrm{h.c.} \right).
\end{align}
We compare the GF spectra obtained from ED for both periodic and open boundary conditions at $t_{12} = 0.25$. 
Under PBC, both methods yield a fully gapped GF zeroes spectrum. Under OBC, the zero-energy edge modes present at $t_{12}=0$ split at finite $t_{12}$, indicating the lifting of edge degeneracy. This demonstrates that GF zeroes capture the trivialization induced by single-particle interchain hopping, which in this case breaks the protecting sublattice symmetry.

\section{BBH model}

In the main text, we focus predominantly  on one-dimensional systems, but the insights gained can be extended to higher dimensional systems. In this section, we consider in particular the Benalcazar–Bernevig–Hughes (BBH) quadrupole insulator model, which hosts a higher-order topological phase protected by chiral and crystalline symmetries~\cite{benalcazar2017quantized}. Its noninteracting limit hosts zero-dimensional corner modes that are the two-dimensional analogs of SSH boundary states in the topological phase. The BBH model is expressed as
\begin{align}
    H_\text{BBH+U} 
        & = \sum\limits_{\mathbf{i}\sigma} \gamma_x\left(c^\dagger_{\mathbf{i}_1,\sigma}c_{\mathbf{i}_3,\sigma}+c^\dagger_{\mathbf{i}_2,\sigma}c_{\mathbf{i}_4,\sigma}+h.c.\right)\\
        & + \gamma_y\left(c^\dagger_{\mathbf{i}_1,\sigma}c_{\mathbf{i}_4,\sigma}-c^\dagger_{\mathbf{i}_2,\sigma}c_{\mathbf{i}_3,\sigma}+h.c.\right) \nonumber \\
        & + \lambda_x\left(c^\dagger_{\mathbf{i}_1,\sigma}c_{\left(\mathbf{i}+\hat{\mathbf{x}}\right)_3,\sigma}+c^\dagger_{\mathbf{i}_4,\sigma}c_{\left(\mathbf{i}+\hat{\mathbf{x}}\right)_2,\sigma}+h.c.\right) \nonumber \\
        & +\lambda_y\left(c^\dagger_{\mathbf{i}_1,\sigma}c_{\left(\mathbf{i}+\hat{\mathbf{y}}\right)_4,\sigma}-c_{\mathbf{i}_3,\sigma}^\dagger c_{\left(\mathbf{i}+\hat{\mathbf{y}}\right)_2,\sigma}+h.c.\right) \nonumber \\
        & + U\sum\limits_{\mathbf{i}} \sum_{\alpha\in\{1, 2, 3, 4\}}\left(\hat{n}_{\mathbf{i}_{\alpha}\uparrow}-\tfrac12\right) \left(\hat{n}_{\mathbf{i}_\alpha\downarrow}-\tfrac12\right). \nonumber
\end{align}

There are four sites per unit cell labeled as $\mathbf{i}_\alpha$ with $\alpha =\{1,2,3,4\}$ located at the following positions $(0,0)$, $(0,1/2)$, $(1/2,0)$ and $(1/2,1/2)$. $\gamma_x$ and $\gamma_y$ are intra-cell hopping along the $x$ and $y$ direction. $\lambda_x$ and $\lambda_y$ are inter-cell hopping along the $x$ and $y$ direction. This can also be understood as a stacking of SSH chain with a $\pi$-flux per unit cell~\cite{benalcazar2017quantized}.

\begin{figure}[h!]
    \centering
    \includegraphics[width=0.7\linewidth]{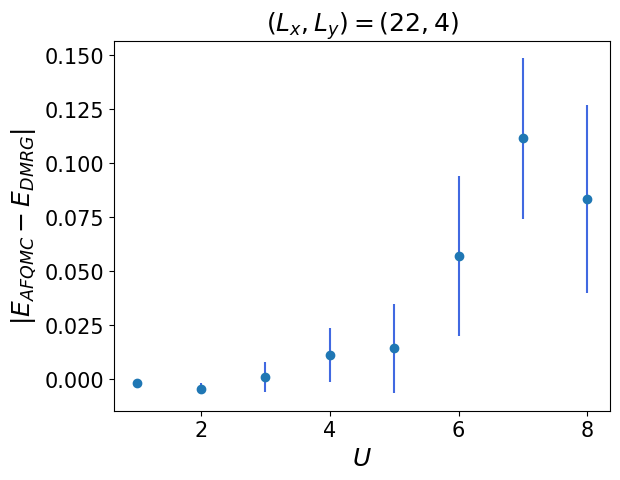}
    \caption{\textbf{Ground-state energy benchmark at half-filling for the BBH model} obtained with DMRG
and AFQMC for a cylinder of size $(L_x, L_y) = (22, 4)$ with OBC with parameters $\lambda_x = \lambda_y = 0.5$ and $\gamma_x = \gamma_y = 1.0$ as a function of $U$.}
    \label{fig:afqmc_vs_dmrg}
\end{figure}

We use AFQMC to study the two-dimensional (2D) limit of this model. In Fig.~\ref{fig:afqmc_vs_dmrg}, we benchmark the half-filled ground-state energy against DMRG for thin cylinders.

\begin{figure}[h!]
    \centering
    \includegraphics[width=\linewidth]{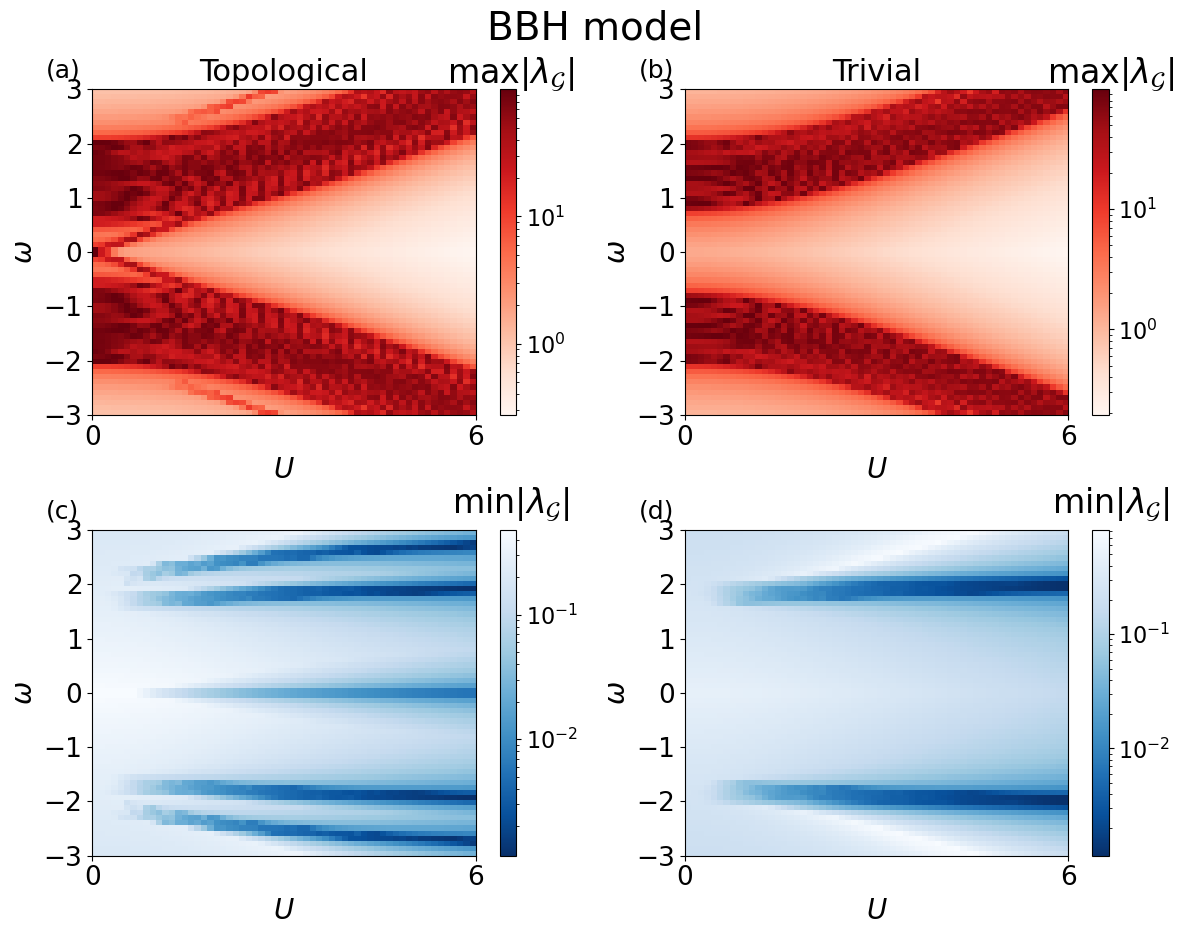}
    \caption{
        \textbf{Poles and zeroes of the topological (left) and trivial (right) BBH model.}
        Maximum (top panels) and minimum (bottom panels) eigenvalues of the interacting Green's function obtained with the Hubbard operator method for the BBH model with open boundary conditions in both direction and length $(L_x,L_y)=(18,18)$ as a function of the Coulomb repulsion $U$. Left : topological BBH model with $\gamma=$~\textonehalf \ and $\lambda=1$. Right : trivial BBH model with $\gamma =1$ and $\lambda=$~\textonehalf.
        }
    \label{fig:poles_zeroes_BBH}
\end{figure}

\begin{figure}[ht!]
    \centering
    \includegraphics[width=\linewidth]{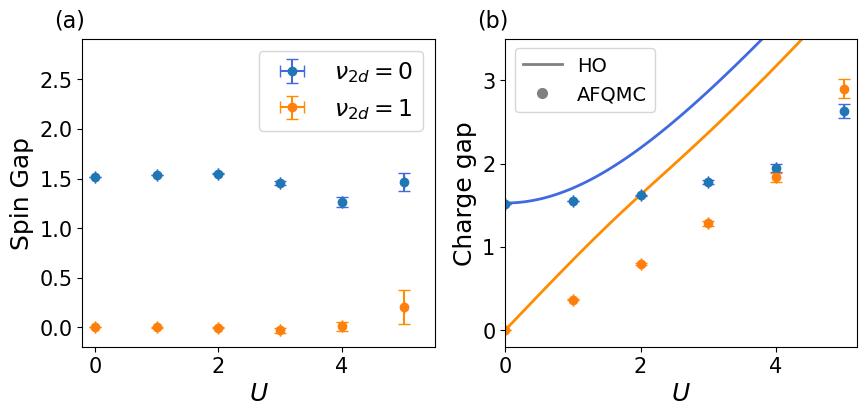}
    \caption{
        \textbf{Spin (a) and charge (b) gaps for a BBH model of size $(L_x,L_y)=(18,18)$ as a function of Coulomb repulsion $U$.} 
        Solid lines represent AFQMC results, whereas dots indicate calculations performed using the Hubbard operator method (HO). $\nu_{2d}=0$ corresponds to the trivial BBH phase (blue, $\gamma =1.0 $, $\lambda = 0.5$), while $\nu_{2d}=1$ corresponds the topological BBH phase (orange $\gamma=0.5$, $\lambda = 0.5$).}
    \label{fig:spin_charge_gaps_BBH}
\end{figure}
In Fig.~\ref{fig:poles_zeroes_BBH}, we compute the GF eigenvalue spectrum $\lambda_\mathcal{G}(\omega,U)$ for the interacting BBH model with OBC in both spatial directions, comparing the trivial and quadrupole-topological phases. As in the SSH case, studying the full GF spectrum rather than only its determinant allows us to track poles and zeroes simultaneously, and to resolve their evolution under interactions. In the topological phase at $U=0$, the boundary response shows four mid-gap corner states, consistent with the quadrupole topology of the BBH model. Once interactions are introduced ($U>0$), these corner modes split into Hubbard-like branches at $\omega=\pm U/2$, signaling that each corner localizes an effective degree of freedom analogous to a Hubbard dot weakly decoupled to the bulk plaquettes. The corresponding GF zeroes appear at the same locations where spectral weight is depleted, forming corner-localized zero modes. Their presence indicates that the interacting system retains the same bulk topological structure as the non-interacting quadrupole insulator, in parallel with the edge-mode splitting mechanism observed in the SSH chain. 

Figure~\ref{fig:spin_charge_gaps_BBH} shows the evolution of the charge and spin gaps as a function of $U$ in both phases obtained through AFQMC. Here we observe only qualitative agreement between the AFQMC and HO methods, in contrast to the SSH chain, where quantitative agreement was obtained between DMRG and HO. This discrepancy can arise from two sources. First, away from half-filling, AFQMC suffers from a sign problem. Calculations can still be performed but they are affected by a bias originating from the constrained-path approximation. Second, the local nature of the holon--doublon basis used in the HO computation may be insufficient to fully capture the two-dimensional correlations of the BBH model, motivating the use of a larger operator basis. Nevertheless, both approaches agree on the topological properties of the phase. \\

The behavior closely mirrors the one-dimensional case: in the topological regime the charge sector becomes immediately gapped for $U>0$, while the spin response remains gapless over the same parameter range, reflecting the persistence of GF zeroes at the corner. In the trivial regime, no corner structure emerges, neither poles nor zeroes form in the mid-gap region, and both charge and spin excitations evolve smoothly with interaction strength.

\end{document}